\documentclass[%
 reprint,
superscriptaddress,
 amsmath,amssymb,
aps,
]{revtex4-2}

\usepackage{graphicx}
\usepackage{dcolumn}
\usepackage{bm}
\usepackage{hyperref}
\usepackage{csquotes}

\usepackage[
margin=0.5in,
]{geometry}

\begin{document}

\preprint{APS/123-QED}

\title{Radial Schmidt mode detector of entangled photons}

\author{Radhika Prasad}
\email{radhikap@iitk.ac.in}
\author{Nilakshi Senapati}
\affiliation{%
	Department of Physics, Indian Institute of Technology Kanpur, Kanpur 208016, India
}
\author{Abhinandan Bhattacharjee}%
\affiliation{Universität Paderborn, Warburger Strasse 100, 33098 Paderborn, Germany}
\author{Anand K Jha}
\email{akjha@iitk.ac.in}
\affiliation{%
 Department of Physics, Indian Institute of Technology Kanpur, Kanpur 208016, India
}%

\date{\today}

\begin{abstract}

High-dimensional spatially entangled two-photon state generated by spontaneous parametric down-conversion process (SPDC) has become a promising resource for several quantum information science applications. For harnessing high-dimensional entanglement advantages, detection capability in the Schmidt basis is a necessity. Spatial entanglement has been explored in several modal bases, such as pixel, azimuthal, and radial modes. Among them, pixel and azimuthal entanglement have been widely utilized due to efficient access to their Schmidt modes, while radial-mode entanglement remains underexploited. This is because for radial coordinates, there is neither a Schmidt-decomposed form for the SPDC photons nor is there a technique for measuring high-dimensional radial Schmidt modes, which is a major roadblock in harnessing radial mode advantages. In this work, we first theoretically show that the azimuthal averaging of SPDC two-photon state yields a radial Schmidt-decomposed form under typical experimental situations. We then demonstrate an innovative approach for extracting the radial Schmidt modes and their spectrum by characterizing the density matrix in the radial basis of one of the SPDC photons. Finally, we report the first-ever measurement of radial Schmidt spectrum of upto 50 radial Schmidt modes with about 98\% fidelity.  

\end{abstract}


\maketitle

\section{Introduction}\label{intro}

High-dimensional entangled two-photon states offer benefits for quantum information applications including increased information capacity \cite{bozinovic2013science}, higher security \cite{cerf2002prl} and error tolerance \cite{bechmann2000prl}, robustness to noise \cite{zhu2021avs}, supersensitive measurements \cite{jha2011praatmol}, and teleportation \cite{luo2019prl}. Among all the potential degrees of freedom available for harnessing the high-dimensional benefits, such as time-energy \cite{zhang2017pra} and pixel entanglement \cite{osullivan2005prl}, the spatial entanglement of the signal and idler photon produced by spontaneous parametric down-conversion (SPDC) is the most suitable for quantum information applications \cite{yao2011advoptphot}. This is because the spatial degree of freedom involves a complete set of discrete and infinite-dimensional modes that are usually propagating solutions to the wave equation, such as the Laguerre-Gaussian $LG_p^l$ modes \cite{allen1992pra}. These modes are characterized by two indices, namely, the azimuthal and the radial. The most often used azimuthal modes are characterised by $e^{il\phi}$ phase dependence and contain $l\hbar$ orbital angular momentum (OAM) per photon, while the radial modes quantify radial dependence of the field. Most works on radial modes have been with the so-called $p$-modes with several potential applications \cite{olvera2023jopt, wang2024optcomm}. The $p$-modes based two-photon entangled states have led to violations of Bell inequalities \cite{zhang2018pra}, uncertainty principle for radial degrees of freedom \cite{zhang2022photres}, and realization of the Einstein-Podolsky-Rosen (EPR) correlations \cite{chen2019prl}.

In the last more than two decades, a lot of effort has gone into generating and detecting high-dimensional entangled two-photon state using SPDC. However, harnessing high-dimensional advantages has been challenging because of the lack of suitable spatial-mode detectors for one-photon field and spatial Schmidt-mode detector for entangled two-photon fields. Schmidt modes are the modes in the basis in which the two-photon state has a Schmidt-decomposed form, and it is very useful for quantifying and accessing high-dimensional entanglement \cite{ekert1995ajp}. An ideal detector for any basis is a mode sorter that spatially separates modes based on their mode index. In the azimuthal basis, the current sorter implementations involve either multiple diffractive elements and suffer from losses that are very difficult to estimate \cite{mirhosseini2013natcomm, berkhout2010prl}, or sort OAM modes with substantial cross-talk \cite{sahu2018optexp}. In the absence of an OAM sorter, the other approach so far has been the projective measurement using a spatial light modulator (SLM) and a single mode fiber (SMF) \cite{mair2001nature, heckenberg1992optlett}, but this approach suffers from mode-dependent losses. More recently, another viable approach has been developed that is based on reconstructing the OAM-mode spectrum by measuring the angular coherence function \cite{jha2011pra, pires2010prl,  kulkarni2017natcomm, kulkarni2018pra}. This has resulted in OAM-mode detectors that work for a broad range of modes and has uniform detection efficiency over the entire range \cite{kulkarni2017natcomm, karan2025sciadv}. 

Unlike azimuthal modes for which efficient mode detectors do exist, currently there is no such detector for radial modes.  The current $p$-mode sorter implementations can sort either only a set of three pre-specified modes \cite{zhou2017prl} or based on whether $l+p$ is odd or even \cite{fu2018optexp}. The efforts based on projective measurements, involving an SLM  and an SMF, have so far yielded a detector that works for only four radial modes with very poor efficiency \cite{choudhary2018optlett} and another detector that have been demonstrated to be efficient for only up to eight modes \cite{bouchard2018optexp}. Thus, we note that the current advancement in radial-mode detection is still at the nascent stage. Furthermore, the reconstruction-based approach which has resulted in near-perfect OAM detectors \cite{jha2011pra, kulkarni2017natcomm, kulkarni2018pra, offer2018commphy, pires2010prl} has so far not been tried for radial modes. 

In this work, we show that the azimuthally-averaged SPDC two-photon state can be approximated as a pure state under typical experimental situations. We thus derive a Schmidt-decomposed form in radial coordinates. Next, we propose a technique for reconstructing the radial-mode spectrum by measuring the azimuthally-averaged radially quasi-homogeneous coherence function and thereby demonstrate an experimental technique for measuring upto 50 radial Schmidt modes with about 98\% fidelity. 

\section{Theory}


The Laguerre-Gaussian (LG) modes, represented as
$LG_p^l(\rho, \phi)$, are exact solutions to the paraxial
Helmholtz equation. The index $l$ measures the OAM of
each photon in the units of $\hbar$, while the index $p$
characterizes the radial variation in the intensity
\cite{allen1992pra}. These modes form a complete basis, and any electric field $E(\rho, \phi)$ can be represented in terms of the projections over them: 
\begin{align}
	E(\bm\rho)=\sum_{l, p} A_{lp} LG_p^l(\bm\rho)=\sum_{l, p} A_{lp} LG_p^l(\rho)e^{il\theta}, \label{field}
\end{align}
where $A_{lp}$ are superposition coefficients, and $\bm{\rho}\equiv \left(\rho,\theta\right)$ is the transverse position vectors in the cylindrical coordinates. In the projective measurement of OAM modes, one essentially measures coefficients $A_{lp}$, quite often with $p=0$ \cite{mair2001nature, heckenberg1992optlett}. In the context of spontaneous parametric down-conversion (SPDC) process, a pump photon at higher frequency down-converts into two photons of lower frequencies, referred to as the signal and idler. The state of the down-converted  photons is given by:
\begin{align}
	|\psi_{si}\rangle =\iint \psi_{si}\left(\bm{\rho}_s, \bm{\rho}_i\right)|\bm\rho_s\rangle |\bm\rho_i\rangle d^2\bm\rho_s d^2\bm\rho_i,  \label{two-photon1}
\end{align}
where under the Gaussian approximation for collinear phase-matching condition, $\psi_{si}\left(\bm{\rho}_s, \bm{\rho}_i\right)$ is given by \cite{schneeloch2016jopt,karan2020jopt,walborn2010phyrep}:
\begin{equation}
	\psi_{si}\left(\bm{\rho}_s, \bm{\rho}_i\right)=A \exp 	\left[-\frac{\left(\bm{\rho}_s+\bm{\rho}_i\right)^2}{4 w_{p}^2}\right] \exp \left[-\frac{\left(\bm{\rho}_s-\bm{\rho}_i\right)^2}{4 \sigma_-^2}\right].\label{psi-spdc}
\end{equation}
Here, $A$ is the normalization constant, and the subscripts $p$, $s$, and $i$ stand for pump, signal and idler, respectively; $w_p$ is the beam-waist width of the pump and $\sigma^{2}_-=\frac{L \lambda_p}{6 \pi}$, where $L$ is the crystal length and $\lambda_p$ is the pump wavelength \cite{schneeloch2016jopt}. For studying the spatial entanglement properties of SPDC photons, the two-photon state $|\psi_{si}\rangle $ is often expressed in the LG basis as \cite{mair2001nature, karan2023prapp}:
\begin{align}
	|\psi_{si}\rangle =\sum_{l_s}\sum_{p_s} \sum_{l_i}\sum_{p_i} C_{l_s,p_s}^{l_i,p_i}|l_s,p_s\rangle_s|l_i,p_i\rangle_i. \label{two-photon2}
\end{align}
where $l_p$ is the OAM mode index of the pump photon, and $|l_s, p_s\rangle_s$ represents the state of the signal photon with indices $l_s$ and $p_s$, etc. Using Eqs.~(\ref{two-photon1}) and (\ref{two-photon2}), we get 
\begin{align}
	C_{l_s,p_s}^{l_i,p_i}=\iint \psi_{si}\left(\bm{\rho}_s, \bm{\rho}_i\right)LG_{p_s}^{l_s*}(\bm{\rho}_s) LG_{p_i}^{l_i*}(\bm{\rho}_i) d^2\bm\rho_s d^2\bm\rho_i.\label{coeff}
\end{align}
We note that Eq.~(\ref{two-photon2}) involves summations over four indices. So, it is not in the Schmidt-decomposed form \cite{ekert1995ajp}, since that should involve summations over only two indices. However, for efficiently harnessing and quantifying entanglement, expressing a state in the Schmidt basis and having the detection capability in that basis is a must.

\subsection{Azimuthal (OAM) Schmidt spectrum of SPDC photons}

We note that for most of the works related to OAM entanglement, one is not interested in radial dependence of the field and that OAM measurements are restricted either to $p_s=p_i=0$ modes \cite{mair2001nature}, or is averaged over all the radial modes \cite{kulkarni2017natcomm, kulkarni2018pra, karan2025sciadv}. For such measurement scenarios, the state of the two-photon field for a Gaussian pump field $(l_p=0)$ is assumed to be \cite{mair2001nature}:
\begin{align}
	|\psi_{si}^{\rm (az)}\rangle=\sum_l \sqrt{C_l}|l\rangle_s |-l\rangle_i,\label{azimuthal-schmidt}
\end{align}
\begin{figure}[t]
	\centering
	\includegraphics[scale=0.65]{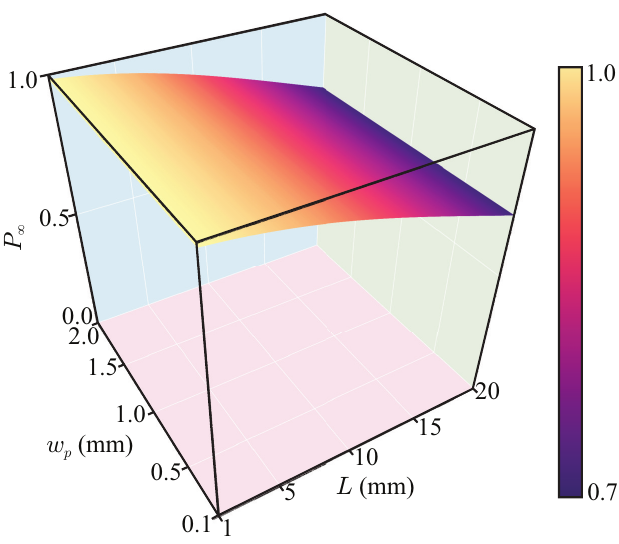}
	\caption{\label{fig:trace} Numerical plot of the purity $P_{\infty}$ of the two-photon state $W_{si}^{(\rm rad)}\left(\rho_s, \rho_i, \rho_{s}^{\prime}, \rho_i^{\prime}\right)$ as a function of crystal length $L$ and beam waist $w_p$. The purity is defined as $P_{\infty}\equiv{\rm Tr}\left[\left[W_{si}^{(\rm rad)}\left(\rho_s, \rho_i, \rho_s^{\prime}, \rho_i^{\prime}\right)\right]^2\right]$  \cite{patoary2019josab}. $P_{\infty}=1$ represents a pure state while $P_{\infty}=0$ represents a completely mixed state. We see that as a function of the beam waist $w_p$, $P_{\infty}$ remains unchanged, whereas it decreases as the crystal length $L$ increases.}
\end{figure}  
which is a Schmidt-decomposed form in the OAM basis. Although this form has not been derived rigorously, it is widely used and has been verified in experiments to a good approximation. In Eq.~(\ref{azimuthal-schmidt}),  $C_l$ is referred to as the OAM Schmidt spectrum, and it represents the probability that the signal and idler photons have OAM $l\hbar$ and $-l\hbar$, respectively. From Eq.~(\ref{azimuthal-schmidt}), the density matrix element corresponding to the signal photon field can be obtained by taking the partial trace over the idler photon:
\begin{align}
	\left\langle \theta_s{\Big|}{\rm Tr}_i\left[{\Big|}\psi_{si}^{\rm (az)}\rangle\langle\psi_{si}^{\rm (az)}{\Big|}\right]{\Big|}\theta_s'\right\rangle=\frac{1}{2\pi}\sum_l C_le^{il(\theta_s-\theta_s')}  
	\equiv W_s^{\rm (az)}(\theta_s, \theta_s').\label{signal-OAM}
\end{align}
The density matrix element $W_s^{\rm (az)}(\theta_s, \theta_s')$ can also referred to as the angular coherence function of the signal field. The above equation is the coherent mode decomposition of the signal field \cite{born&wolf1999, jha2011pra}. Since $C_l$ is the weightage of the OAM modes, it is referred to as the OAM spectrum. Thus, from Eqs.~(\ref{azimuthal-schmidt}) and (\ref{signal-OAM}), we have that the OAM spectrum of one of the down-converted photons is same as the OAM Schmidt spectrum of the entangled two-photon field \cite{jha2011pra}---this fact is used for experimentally measuring the OAM Schmidt spectrum \cite{pires2010prl, kulkarni2017natcomm, kulkarni2018pra, karan2025sciadv}. 

\begin{figure*}[t]
	\centering
	\includegraphics[width =\textwidth]{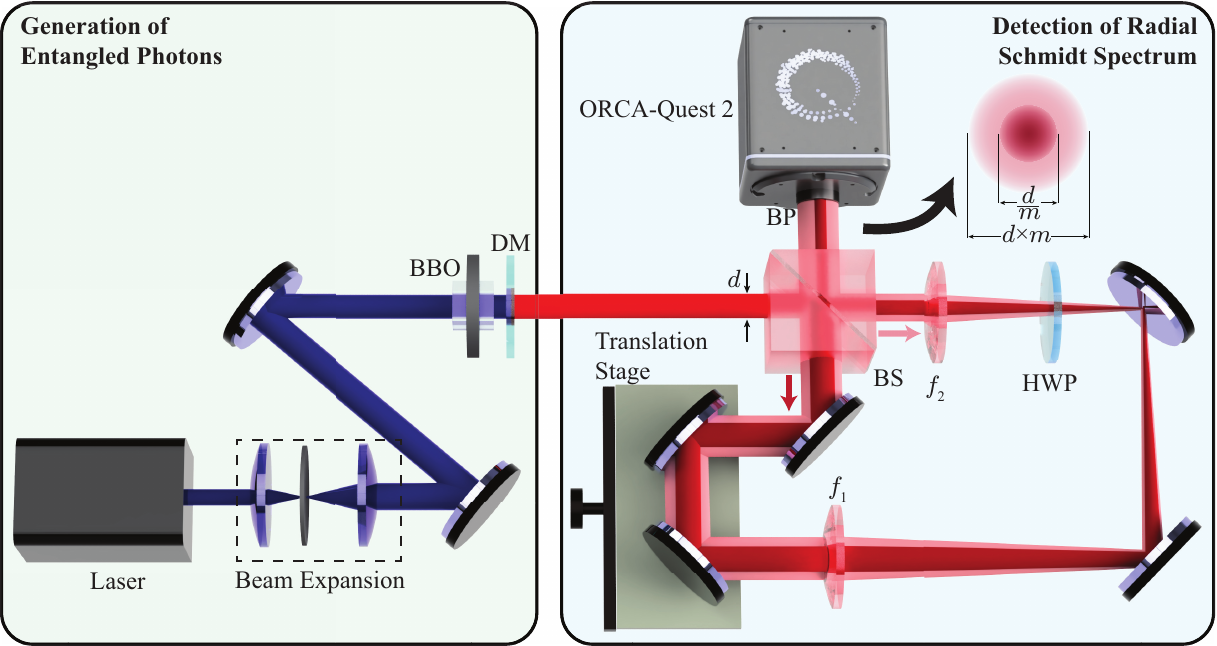}
	\caption{\label{fig:setup}Experimental setup involving common-path Sagnac interferometer. Laser: ultraviolet (UV) laser; BS: beam splitter; BBO: $\beta$-Barium Borate crystal; HWP: half-wave plate (optic axis orientation $\phi$ kept at $0$ and $\pi/4$ radians from the horizontal); DM: dichroic mirror; BP: bandpass filter. We represent the reflected and transmitted beams at the BS with different colours (dark red and light red, respectively) to highlight the magnification due to $4f$ lens configuration. The BBO crystal is used for generation of entangled photons and the interferometer detects the radial Schmidt spectrum.}
\end{figure*}

\subsection{Radial Schmidt spectrum of SPDC photons}

\begin{figure*}[t]
	\centering
	\includegraphics[width =\textwidth]{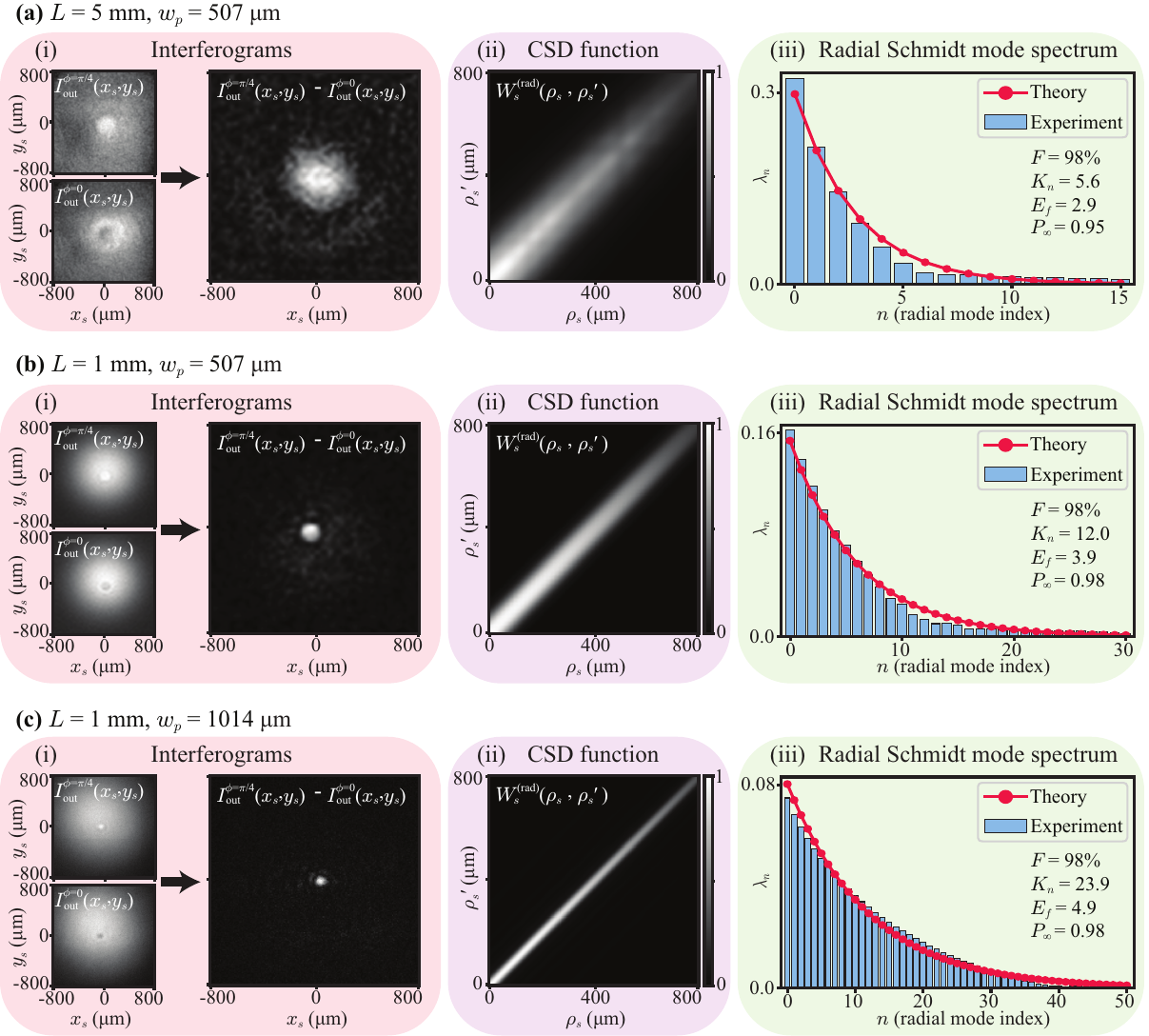}
	\caption{\label{fig:results}Experimental results: Radial Schmidt mode spectrum of entangled photons for different combinations of $L$ and $w_p$. (a) $L=5$ mm and $w_p=507 \mu$m; (b) $L=1$ mm and $w_p=507 \mu$m; (c) $L=1$ mm and $w_p=1014 \mu$m. (i) Recorded interferograms at $\phi=\pi/4$ and $\phi=0$, and their corresponding difference. (ii) The signal photon radial cross-spectral density (CSD) function $W_s^{(\rm rad)} \left(\rho_s, \rho_s^{\prime}\right)$. (iii) Radial Schmidt mode spectrum obtained by finding eigenvalues of $W_s^{(\rm rad)} \left(\rho_s, \rho_s^{\prime}\right)$. Red curve is the theoretical prediction and blue bars are the experimentally obtained values. $F$ is the fidelity, $K_n$ is the Schmidt number, $E_f$ is the entanglement of formation, $P_{\infty}$ is the two-photon state purity.} \end{figure*}

Next, we work out an analogous technique for measuring the radial Schmidt spectrum of entangled two-photon fields. In the radial coordinates, the measurement techniques have so far demonstrated measuring radial modes only as $p$-modes, and up to $p=8$ for single photon states \cite{zhou2017prl, fu2018optexp, choudhary2018optlett, bouchard2018optexp}.  Although more recently, measurements of radial modes of entangled photons have also been demonstrated, through direct p-mode projective measurements \cite{zhang2018pra} and through interferometric reconstruction of the spatial biphoton state, from which the p-mode spectrum was subsequently extracted \cite{zia2023natphot}, to the best of our knowledge, the Schmidt-decomposed form in radial coordinates has neither been derived analytically, nor is there any experimental technique for measuring it.
%
%
%
%
To represent two-photon state in terms of only radial variables $(\rho_s,\rho_i)$, we take partial trace of $|\psi_{si}\rangle$ in Eq. \ref{two-photon1} over the azimuthal variables $(\theta_s,\theta_i)$. This yields a two-photon state, the density matrix element of which can be written as  
\begin{equation} \label{eq:w si rad}
	W_{si}^{(\rm rad)}\left(\rho_s, \rho_i, \rho_{s}^{\prime}, \rho_i^{\prime}\right)=\iint \psi_{si}^*\left(\boldsymbol{\rho}_s^{\prime}, \boldsymbol{\rho}_i^{\prime}\right) \psi_{s i}\left(\boldsymbol{\rho}_s, \boldsymbol{\rho}_i\right) d \theta_s d \theta_i.
\end{equation}	
$W_{si}^{(\rm rad)}\left(\rho_s, \rho_i, \rho_{s}^{\prime}, \rho_i^{\prime}\right)$ can also be referred to as the two-photon radial cross-spectral density function \cite{jha2010pra}. We take the state to be normalized, that is, ${\rm Tr}\left[\left[W_{si}^{(\rm rad)}\left(\rho_s, \rho_i, \rho_{s}^{\prime}, \rho_i^{\prime}\right)\right]\right]=1$, where $[\cdots]$ denotes matrix representation. We note that the Schmidt decomposition for a two-photon state exists only if the state is pure and that the state represented by Eq.~(\ref{eq:w si rad}) is, strictly speaking, a mixed state. However, if the mixedness of this state is negligible or in other words if the purity of the state is close to unity, we can take the state to be pure, which will then have a Schmidt decomposition. A generic quantifier of the purity of any quantum state is the intrinsic degree of coherence, represented by $P_\infty$ for continuous variables bases, such as position, momentum, etc. $P_\infty$ is defined as $P_\infty={\rm Tr}[\rho^2]$, where $\rho$ is the density matrix of the quantum state \cite{patoary2019josab, meher2020josab}. And therefore, for the above state we have,  $P_\infty={\rm Tr}[\rho^2]=\left[\left[W_{si}^{(\rm rad)}\left(\rho_s, \rho_i, \rho_s^{\prime}, \rho_i^{\prime}\right)\right]^2\right]$. $P_{\infty}$ ranges from 0 to 1, with $P_{\infty}=1$ representing a pure state and $P=0$ representing the completely mixed state. Using Eqs.~(\ref{psi-spdc}) and (\ref{eq:w si rad}), we evaluate $P_\infty$ for various values of the crystal length $L$ and the pump beam waist $w_p$ for $\lambda_p=355 nm$. Figure \ref{fig:trace} plots $P_{\infty}$ as a function of $L$ and $w_p$. We find that as a function of $w_p$, $P_{\infty}$ remains unchanged, whereas $P_{\infty}$ increases as $L$ decreases. Thus, we see that as the crystal becomes thinner, $P_{\infty}$ approaches unity. In our experiments reported below, we employ crystals with $L=1$ mm and $L=5$ mm, for which the $P_{\infty}$ comes out to be $0.95$ and $0.98$, respectively. More importantly, as we show in the experimental section, for producing high-dimensional radial Schmidt spectrum, one requires thinner crystal. Therefore, in the case of higher-dimensional Schmidt spectrum, the accuracy of this approximation becomes progressively better. Thus for such crystal lengths, we  approximate $P_{\infty}$ to be unity and take the state represented by Eq.~(\ref{eq:w si rad}) to be pure,  and we therefore write  the density matrix element of Eq.~(\ref{eq:w si rad}) as  $W_{si}^{(\rm rad)}\left(\rho_s, \rho_i, \rho_{s}^{\prime}, \rho_i^{\prime}\right)\approx \psi_{si}^{\rm * (rad)}\left(\rho_s, \rho_i\right)\psi_{s i}^{\rm (rad)}\left(\rho_s', \rho_i'\right)$, where $\psi_{s i}^{\rm (rad)}\left(\rho_s, \rho_i\right)$ represents a pure two-photon wavefunction, which must have a Schmidt-decomposed form that can be represented as
\begin{figure*}[t]
	\centering
	\includegraphics[width =\textwidth]{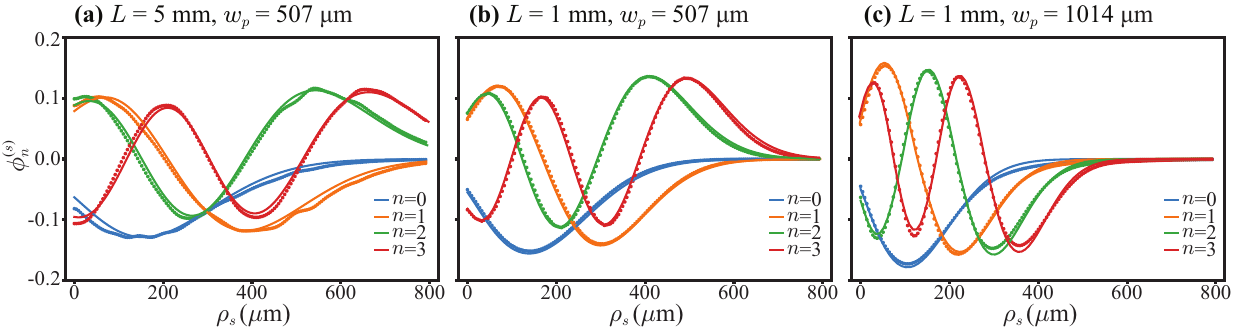}
	\caption{\label{fig:eigenfunctions} Schmidt modes for the three different experimental conditions: (a) $L=5$ mm and $w_p=507 \mu$m; (b) $L=1$ mm and $w_p=507 \mu$m; (c) $L=1$ mm and $w_p=1014 \mu$m. The dotted lines are experimental measurements, while the solid lines are the theoretical predictions.} \end{figure*}
\begin{equation}\label{eq:psi_schmidt-wave}
	\psi_{s i}^{\rm(rad)}\left(\rho_s, \rho_i\right)=\sum_n \sqrt{\lambda_n} \phi_n^{(s)}\left(\rho_s\right) \phi_n^{(i)}\left(\rho_i\right).
\end{equation}
Here, $\lambda_n$ can be referred to as the radial Schmidt spectrum and $\phi_n^{(s)}\left(\rho_s\right)$ as the radial Schmidt modes. We take the SPDC phase matching to be low-gain, degenerate, and Type-I collinear, in which case the eigenfunctions  $\phi_n^{(s)}\left(\rho_s\right)$ and $\phi_n^{(i)}\left(\rho_i\right)$ will have the same functional forms. Equation (\ref{eq:psi_schmidt-wave}) can be referred to as the radial Schmidt decomposition of the entangled two-photon field, and it is analogous to the azimuthal Schmidt decomposition given in Eq.~(\ref{azimuthal-schmidt}). From the two-photon state in Eq.~(\ref{eq:psi_schmidt-wave}), the state of the signal photon can be calculated by taking the partial trace over the idler. Thus, the density matrix element corresponding to the signal photon in the radial basis can be shown to be:
\begin{equation}\label{eq:w_schmidt}
	\begin{aligned}
		W_s^{(\rm rad)} \left(\rho_s, \rho_s^{\prime}\right) & =\int \psi_{s i}^{*\rm (rad)}\left(\rho_s^{\prime}, \rho_i\right) \psi_{s i}^{\rm (rad)}\left(\rho_s, \rho_i\right) d \rho_i \\
		& =\sum_n \lambda_n \phi_n^{*(s)}\left(\rho_s^{\prime}\right) \phi_n^{(s)}\left(\rho_s\right).
	\end{aligned}
\end{equation}
We note that $W_s^{(\rm rad)} \left(\rho_s, \rho_s^{\prime}\right)$ can be referred to as the  radial cross-spectral density function corresponding to the signal photon, and it quantifies coherence in the radial coordinates \cite{bhattacharjee2023josaa}. Furthermore, $W_s^{(\rm rad)} \left(\rho_s, \rho_s^{\prime}\right)$ in Eq.~(\ref{eq:w_schmidt}) is the coherent mode representation \cite{prasad2025optexp, bhattacharjee2023josaa}, and therefore, $\lambda_n$ in the above equation represents the radial-mode spectrum of signal photon, just as $C_l$ represents OAM-mode spectrum in Eq.~(\ref{signal-OAM}). Comparing Eqs.~(\ref{eq:psi_schmidt-wave}) and (\ref{eq:w_schmidt}), we find that the radial Schmidt spectrum of entangled two-photon field is same as the radial-mode spectrum of either the signal or the idler photon. Equation (\ref{eq:w_schmidt}) is the main theoretical result of this work, which implies that by measuring the radial cross-spectral density function $W_s^{(\rm rad)} \left(\rho_s, \rho_s^{\prime}\right)$ and then finding its coherent mode representation, one can obtain the radial Schmidt spectrum $\lambda_n$ and the corresponding Schmidt modes $\phi_n(\rho_s)$ of the entangled two-photon states. We note that, unlike Eq.~(\ref{azimuthal-schmidt}), in which case the Schmidt modes are assumed to be the OAM modes, the functional form of the Schmidt modes has not been assumed in Eq.~(\ref{eq:w_schmidt}) but is obtained through the coherent mode represented of the experimentally measured $W_s^{(\rm rad)} \left(\rho_s, \rho_s^{\prime}\right)$. Thus, our method works for any radial-mode bases and not just the $p$-modes which has been the subject of investigation in most prior works \cite{zhou2017prl, fu2018optexp, choudhary2018optlett, bouchard2018optexp}.

\section{Experimental Measurement Scheme}

Figure \ref{fig:setup} shows the experimental setup, involving a Sagnac-type interferometer. The interferometer measures any cross-spectral density function $W_{s}\left(\boldsymbol{\rho}_s, \boldsymbol{\rho}_s^{\prime}\right)$ that is quasi-homogeneous \cite{prasad2025optexp}, or in other words, any $W_{s}\left(\boldsymbol{\rho}_s, \boldsymbol{\rho}_s^{\prime}\right)$ that can be expressed as
\begin{equation}
	W(\bm{\rho}_s,\bm{\rho'}_s) = \sqrt{I(\bm{\rho}_s)I(\bm{\rho'}_s)}\mu(\bm{\rho}_s-\bm{\rho'}_s).
	\label{W}
\end{equation}
Here, $I(\bm{\rho}_s)=W(\bm{\rho}_s,\bm{\rho}_s)$ is the transverse position probability of the signal photon and $\mu(\bm{\rho}_s-\bm{\rho'}_s)$ represents the degree of spatial coherence between $\bm{\rho}_s$ and $\bm{\rho'}_s$. The quasi-homogeneity condition requires that (i) the width of $\mu(\bm{\rho}_s-\bm{\rho'}_s)$ must be much smaller than the width of $I(\bm{\rho}_s)$ and (ii) $\mu(\bm{\rho}_s-\bm{\rho'}_s)$ should depend only on the difference coordinate $\bm{\rho}_s-\bm{\rho'}_s$. Given the two-photon wavefunction in Eq.~(\ref{two-photon1}), it can be numerically shown that for type-I, collinear, degenerate SPDC process for which $\sigma_{-} \ll w_{p}$ \cite{karan2020jopt}, the functional form of $W(\bm{\rho}_s,\bm{\rho'}_s)$ is that of a radial Gaussian Schell-model (RGSM) beam \cite{prasad2025optexp,bhattacharjee2023josaa}, which is an example of a quasi-homogeneous field.

The interferometer of Fig.~\ref{fig:setup} consists of $4f$ lens configuration for magnification and a half-wave plate (HWP) for phase control. The field gets magnified by a factor of $m=f_1/f_2$ in alternative 1 (light red, transmitted twice at the beam splitter) and demagnified by the same factor $m$ in alternative 2 (dark red, reflected twice at the beam splitter). The detection probability $I_{\text {out }}^\phi\left(\bm\rho_s\right)$ at the camera plane is given by (for details see Ref.~\cite{prasad2025optexp}): 
\begin{align}
	I_{\text {out }}^\phi\left(\bm\rho_s \right)=I_{\text {out }}^\phi\left(\rho_s, \theta_s \right)= I\left(m\rho_s, \theta_s \right) + I\left(\frac{\rho_s}{m}, \theta_s \right) -2W\left(m\rho_s, \theta_s; \frac{\rho_s}{m}, \theta_s \right) \cos (4 \phi).
\end{align}
where $\phi$ is the half-wave plate (HWP) angle with respect to the horizontal axis. Here, we have taken the beam splitter to be 50:50 and the cross-spectral density function $W\left(m\rho_s, \theta_s; \frac{\rho_s}{m}, \theta_s \right)$ to be real. $I\left(m\rho_s, \theta_s \right)$ and  $I\left(\frac{\rho_s}{m}, \theta_s \right) $ are the detection probabilities of the individual interfering paths. We note that $W\left(m\rho_s, \theta_s; \frac{\rho_s}{m}, \theta_s \right))$ can be measured by recording $I_{\text {out }}^\phi\left(\rho_s, \theta_s \right)$ at $\phi=0$ and $\phi=\pi/4$, that is, $W\left(m\rho_s, \theta_s; \frac{\rho_s}{m}, \theta_s \right)) = \left[I_{\text {out }}^{\phi=\pi / 4}(\rho_s, \theta_s)-I_{\text {out }}^{\phi=0}(\rho_s, \theta_s)\right]$, and thereby the radial cross-spectral density function of the signal photon can be obtained as
\begin{align}
	W_s^{(\rm rad)} \left(m\rho_s, \frac{\rho_s}{m}\right)=\frac{1}{2\pi}\int_\pi^\pi
	W\left(m\rho_s, \theta_s; \frac{\rho_s}{m}, \theta_s \right)) d\theta_s.
\end{align}
For measuring $I\left(m \rho_s \right)$ and $ I\left(\frac{\rho_s}{m}\right)$, one needs to remove the beam splitter from the interferometer. This way, field from only one interfering paths reaches the camera and thus one measures $I\left(\frac{\rho_s}{m}\right)$. Since it is a common path interferometer, one cannot directly measure $I\left(m\rho_s \right)$, but it can be assessed by rescaling the measured $I\left(\frac{\rho_s}{m}\right)$ by the factor $m$. The degree of coherence function can therefore be obtained as 
\begin{align}
	\mu^{(\rm rad)}_s\left(m{\rho}_s, \frac{{\rho}_s}{m} \right) = \frac{W_s^{(\rm rad)} \left(m{\rho}_s, \frac{{\rho}_s}{m} \right)}{\sqrt{I(m {\rho}_s)I(\frac{{\rho}_s}{m})}}. \label{radial-mu}
\end{align}
Now, by utilizing the quasi-homogeneity of the field, we write: $\mu^{(\rm rad)}_s\left(m{\rho}_s, \frac{{\rho}_s}{m} \right) =\mu^{(\rm rad)}_s\left(m{\rho}_s -\frac{{\rho}_s}{m} \right)=\mu^{(\rm rad)}_s\left[\rho_s\left(m-\frac{1}{m}\right)\right]$. Thus, from the experimental measurements, one can obtain $\mu^{(\rm rad)}_s\left(m{\rho}_s-\frac{{\rho}_s}{m} \right)$ for a given value of $m$. We note that the degree of coherence function obtained this way depends only on one variable, $\rho_s\left(m-\frac{1}{m}\right)$. However, the degree of coherence function is a two-point correlation function. So, from the measured values of $\mu^{(\rm rad)}_s\left[\rho_s\left(m-\frac{1}{m}\right)\right]$, one can obtain the two-point degree of coherence function $\mu^{(\rm rad)}_s(\rho, \rho')$ by realizing that the value of $\mu^{(\rm rad)}_s(\rho, \rho')$ for each pair of $(\rho, \rho')$ is related to the single-variable $\mu^{(\rm rad)}_s(\rho-\rho')$ by $\mu^{(\rm rad)}_s(\rho, \rho')=\mu^{(\rm rad)}_s(\rho-\rho')$. Once $\mu^{(\rm rad)}_s(\rho, \rho')$ is numerically reconstructed, one can numerically obtain the cross-spectral density function $W^{(\rm rad)}_s(\rho, \rho')$ using Eq.~(\ref{radial-mu}). By numerically diagonalizing $W^{(\rm rad)}_s(\rho, \rho')$, one can finally obtain the radial coherent modes and radial mode spectrum of Eq.~(\ref{eq:w_schmidt}) and thus the radial Schmidt modes of Eq.~(\ref{eq:psi_schmidt-wave}).

\section{Experimental Setup}

Figure \ref{fig:setup} shows the experimental setup. A continuous wave ultraviolet (UV) laser of wavelength 355 nm pumps a type-I $\beta$-barium borate (BBO) crystal to generate collinear degenerate SPDC photons centered at wavelength 710 nm. We perform experiments with two different pump beam waists, $w_p=$507 $\mu$m and $1014$. We use a lens combination after the pump laser to achieve the desired pump beam waist at the crystal plane. A dichroic mirror (DM) is kept just after the crystal to block the UV pump and transmit the signal and idler photons. In our experiments, we use two different lengths of the crystal, $L=5$ mm and $L=1$ mm, for which $P_\infty$ comes out to be 0.95 and 0.98, such that the two-photon state in Eq.~(\ref{eq:w si rad}) can be approximated to be pure. We note that the signal and idler fields generated by SPDC co-propagate through the interferometer. The interferogram recorded by the camera is the sum of the intensity distributions of the signal field and idler field at the camera plane. For type-I degenerate SPDC, signal and idler fields are spatially identical, that is, both the signal and idler fields have identical intensity distribution at the camera plane. Therefore, the recorded interferogram is equivalent to that produced by either signal, or idler photon field. The down-converted field enters the interferometer shown in Fig.~\ref{fig:setup} and the interference pattern is recorded by an ORCA-Quest 2 qCMOS camera. For the $4f$ imaging system inside the interferometer, we use lenses with focal lengths $f_1=400$ mm and $f_2=200$ mm, and thus in our experiments we have $m=2$. A translation stage is introduced in to ensure that separation between the two lenses remains intact. A 10 nm bandpass filter centred at 710 nm is place in front of the camera to ensure high temporal coherence of the detected field.

\section{Results and Discussions}

Figure~\ref{fig:results} shows the experimental data and the retrieved Schmidt spectra for three different setting of $w_p$ and $L$. Fig. \ref{fig:results}(a) corresponds to $w_p=507 \mu$m and $L=5$ mm; Fig. \ref{fig:results}(b) corresponds to $w_p=507 \mu$m and $L=1$ mm; Fig. \ref{fig:results}(c) corresponds to $w_p=1014 \mu$m and $L=1$ mm. For each setting, the sub-figure (i) shows  the recorded interferograms at $\phi=\pi/4$ (constructive interference) and $\phi=0$ (destructive interference), along with their differences; the sub-figure (ii) presents the reconstructed radial cross-spectral density function $W^{(\rm rad)}_s(\rho, \rho')$; and the sub-figure (iii) presents the normalized radial Schmidt spectrum of the entangled two-photon state. Along with the reconstructed radial Schmidt spectrum indicated by blue histogram, the sub-figures (iii) shows the theoretical predictions plotted in red. We find an excellent match between theory and experiment with a fidelity of 98\% for all the reconstructed Schimdt spectra. The fidelity $F$ characterizes the accuracy of measurement, and is calculated as the similarity between the theoretical prediction and experimental measurement \cite{karan2023prapp}. We note that the diagonal width of $W^{(\rm rad)}_s(\rho, \rho')$ corresponds to the radial coherence width while the anti-diagonal width $W^{(\rm rad)}_s(\rho, \rho')$ is the radial width of the intensity of the field. We note that the diagonal width of $W^{(\rm rad)}_s(\rho, \rho')$ decreases with increasing $w_p$ and decreasing $L$. And as $W^{(\rm rad)}_s(\rho, \rho')$ decreases, the radial Schmidt spectrum becomes wider and the dimensionality of the state increases. We quantify the dimensionality of the state by evaluating the Schmidt number $K_n=1 / \sum_n \lambda_n^2$, which provides the effective number of radial modes contained by the SPDC state. We find that $K_n$ increases from $5.6$ to $23.9$ as $w_p$ increases and $L$ decreases. This increasing trend in $K_n$ is a qualitative signature of increase of radial entanglement. This trend is also captured by entanglement of formation $E_f$ \cite{schneeloch2019natcomm,prasad2024prapp}, another entanglement certifier. A positive value of $E_f$ implies the presence of entanglement \cite{prevedel2011natphy}. In our case, $E_f$ can be evaluated as $E_f=-\sum_n \lambda_n \log _2 \lambda_n$ \cite{bennett1996pra}. The values of $K_n$ and $E_f$ are also shown in column Fig. \ref{fig:results}(iii). The radial Schmidt modes corresponding to the three different setting of $w_p$ and $L$ are plotted in Figure~\ref{fig:eigenfunctions}, which shows the first four mode profiles (dots) along side their theoretical predictions (solid curve). We again see a  very good match between the theoretical and experimental plots. Finally, we note that since our scheme is based on Sagnac interferometer, which is a common-path interferometer and is known to not suffer from the typical interferometric stability issues, it is quite robust and thus works for high-dimensional states.

\section{Conclusion}

In this work, we have shown that the azimuthally averaged SPDC two-photon state can be well approximated as a radially pure two-photon state under typical experimental conditions, which allows us to derive a Schmidt-decomposed representation in radial coordinates for the entangled two-photon fields produced by SPDC under low-gain, degenerate, collinear Type-I phase-matching condition. We have then proposed and demonstrated a scheme for reconstructing the radial Schmidt spectrum and modes by directly characterizing the density matrix in the radial basis of one of the SPDC photons. This method has enabled the measurements of up to 50 radial Schmidt modes with fidelity 98\%. To the best of our knowledge, this represents atleast five-fold improvement in the detection capability of radial modes over the existing radial-mode detectors. In contrast to most Schmidt-mode characterization schemes, our approach does not require coincidence measurements, which makes our approach accurate and resource efficient. Furthermore, the existing methods works only if the radial Schmidt basis is known a-proiri whereas in our scheme the radial Schmidt basis is directly measured. Additionally, the technique is readily applicable to the characterization of the Schmidt spectrum of bright SPDC sources well beyond the photon-pair regime, which can estimate squeezing of independent radial Schmidt modes. We therefore expect our method to have important implications for high-dimensional quantum information application.

\vspace{1em}

\paragraph*{Data Availability:} Data underlying the results presented in this paper are not publicly available at this time but may be obtained from the authors upon reasonable request.

\vspace{1em}

\paragraph*{Acknowledgements:} R.P and N.S. thank the Prime Minister’s Research Fellowship (PMRF), Government of India, for financial support.

\vspace{1em}

\paragraph*{Funding:} The Science and Engineering Research Board through grants STR/2021/000035 and CRG/2022/003070. The Department of Science and Technology, Government of India, through grants DST/ICPS/QuST/Theme-1/2019 and the National Quantum Mission (NQM).


\bibliography{radial_spdc_optquant}

\begin{thebibliography}{46}%
\makeatletter
\providecommand \@ifxundefined [1]{%
 \@ifx{#1\undefined}
}%
\providecommand \@ifnum [1]{%
 \ifnum #1\expandafter \@firstoftwo
 \else \expandafter \@secondoftwo
 \fi
}%
\providecommand \@ifx [1]{%
 \ifx #1\expandafter \@firstoftwo
 \else \expandafter \@secondoftwo
 \fi
}%
\providecommand \natexlab [1]{#1}%
\providecommand \enquote  [1]{``#1''}%
\providecommand \bibnamefont  [1]{#1}%
\providecommand \bibfnamefont [1]{#1}%
\providecommand \citenamefont [1]{#1}%
\providecommand \href@noop [0]{\@secondoftwo}%
\providecommand \href [0]{\begingroup \@sanitize@url \@href}%
\providecommand \@href[1]{\@@startlink{#1}\@@href}%
\providecommand \@@href[1]{\endgroup#1\@@endlink}%
\providecommand \@sanitize@url [0]{\catcode `\\12\catcode `\$12\catcode
  `\&12\catcode `\#12\catcode `\^12\catcode `\_12\catcode `\%12\relax}%
\providecommand \@@startlink[1]{}%
\providecommand \@@endlink[0]{}%
\providecommand \url  [0]{\begingroup\@sanitize@url \@url }%
\providecommand \@url [1]{\endgroup\@href {#1}{\urlprefix }}%
\providecommand \urlprefix  [0]{URL }%
\providecommand \Eprint [0]{\href }%
\providecommand \doibase [0]{https://doi.org/}%
\providecommand \selectlanguage [0]{\@gobble}%
\providecommand \bibinfo  [0]{\@secondoftwo}%
\providecommand \bibfield  [0]{\@secondoftwo}%
\providecommand \translation [1]{[#1]}%
\providecommand \BibitemOpen [0]{}%
\providecommand \bibitemStop [0]{}%
\providecommand \bibitemNoStop [0]{.\EOS\space}%
\providecommand \EOS [0]{\spacefactor3000\relax}%
\providecommand \BibitemShut  [1]{\csname bibitem#1\endcsname}%
\let\auto@bib@innerbib\@empty
\bibitem [{\citenamefont {Bozinovic}\ \emph {et~al.}(2013)\citenamefont
  {Bozinovic}, \citenamefont {Yue}, \citenamefont {Ren}, \citenamefont {Tur},
  \citenamefont {Kristensen}, \citenamefont {Huang}, \citenamefont {Willner},\
  and\ \citenamefont {Ramachandran}}]{bozinovic2013science}%
  \BibitemOpen
  \bibfield  {author} {\bibinfo {author} {\bibfnamefont {N.}~\bibnamefont
  {Bozinovic}}, \bibinfo {author} {\bibfnamefont {Y.}~\bibnamefont {Yue}},
  \bibinfo {author} {\bibfnamefont {Y.}~\bibnamefont {Ren}}, \bibinfo {author}
  {\bibfnamefont {M.}~\bibnamefont {Tur}}, \bibinfo {author} {\bibfnamefont
  {P.}~\bibnamefont {Kristensen}}, \bibinfo {author} {\bibfnamefont
  {H.}~\bibnamefont {Huang}}, \bibinfo {author} {\bibfnamefont {A.~E.}\
  \bibnamefont {Willner}},\ and\ \bibinfo {author} {\bibfnamefont
  {S.}~\bibnamefont {Ramachandran}},\ }\bibfield  {title} {\bibinfo {title}
  {Terabit-scale orbital angular momentum mode division multiplexing in
  fibers},\ }\href {https://doi.org/10.1126/science.1237861} {\bibfield
  {journal} {\bibinfo  {journal} {Science}\ }\textbf {\bibinfo {volume}
  {340}},\ \bibinfo {pages} {1545} (\bibinfo {year} {2013})}\BibitemShut
  {NoStop}%
\bibitem [{\citenamefont {Cerf}\ \emph {et~al.}(2002)\citenamefont {Cerf},
  \citenamefont {Bourennane}, \citenamefont {Karlsson},\ and\ \citenamefont
  {Gisin}}]{cerf2002prl}%
  \BibitemOpen
  \bibfield  {author} {\bibinfo {author} {\bibfnamefont {N.~J.}\ \bibnamefont
  {Cerf}}, \bibinfo {author} {\bibfnamefont {M.}~\bibnamefont {Bourennane}},
  \bibinfo {author} {\bibfnamefont {A.}~\bibnamefont {Karlsson}},\ and\
  \bibinfo {author} {\bibfnamefont {N.}~\bibnamefont {Gisin}},\ }\bibfield
  {title} {\bibinfo {title} {Security of quantum key distribution using d-level
  systems},\ }\href {https://doi.org/10.1103/PhysRevLett.88.127902} {\bibfield
  {journal} {\bibinfo  {journal} {Phys. Rev. Lett.}\ }\textbf {\bibinfo
  {volume} {88}},\ \bibinfo {pages} {127902} (\bibinfo {year}
  {2002})}\BibitemShut {NoStop}%
\bibitem [{\citenamefont {Bechmann-Pasquinucci}\ and\ \citenamefont
  {Peres}(2000)}]{bechmann2000prl}%
  \BibitemOpen
  \bibfield  {author} {\bibinfo {author} {\bibfnamefont {H.}~\bibnamefont
  {Bechmann-Pasquinucci}}\ and\ \bibinfo {author} {\bibfnamefont
  {A.}~\bibnamefont {Peres}},\ }\bibfield  {title} {\bibinfo {title} {Quantum
  cryptography with 3-state systems},\ }\href
  {https://doi.org/10.1103/PhysRevLett.85.3313} {\bibfield  {journal} {\bibinfo
   {journal} {Physical Review Letters}\ }\textbf {\bibinfo {volume} {85}},\
  \bibinfo {pages} {3313} (\bibinfo {year} {2000})}\BibitemShut {NoStop}%
\bibitem [{\citenamefont {Zhu}\ \emph {et~al.}(2021)\citenamefont {Zhu},
  \citenamefont {Tyler}, \citenamefont {Valencia}, \citenamefont {Malik},\ and\
  \citenamefont {Leach}}]{zhu2021avs}%
  \BibitemOpen
  \bibfield  {author} {\bibinfo {author} {\bibfnamefont {F.}~\bibnamefont
  {Zhu}}, \bibinfo {author} {\bibfnamefont {M.}~\bibnamefont {Tyler}}, \bibinfo
  {author} {\bibfnamefont {N.~H.}\ \bibnamefont {Valencia}}, \bibinfo {author}
  {\bibfnamefont {M.}~\bibnamefont {Malik}},\ and\ \bibinfo {author}
  {\bibfnamefont {J.}~\bibnamefont {Leach}},\ }\bibfield  {title} {\bibinfo
  {title} {Is high-dimensional photonic entanglement robust to noise?},\
  }\bibfield  {journal} {\bibinfo  {journal} {AVS Quantum Science}\ }\textbf
  {\bibinfo {volume} {3}},\ \href {https://doi.org/10.1116/5.0033889}
  {10.1116/5.0033889} (\bibinfo {year} {2021})\BibitemShut {NoStop}%
\bibitem [{\citenamefont {Jha}\ \emph {et~al.}(2011{\natexlab{a}})\citenamefont
  {Jha}, \citenamefont {Agarwal},\ and\ \citenamefont
  {Boyd}}]{jha2011praatmol}%
  \BibitemOpen
  \bibfield  {author} {\bibinfo {author} {\bibfnamefont {A.~K.}\ \bibnamefont
  {Jha}}, \bibinfo {author} {\bibfnamefont {G.~S.}\ \bibnamefont {Agarwal}},\
  and\ \bibinfo {author} {\bibfnamefont {R.~W.}\ \bibnamefont {Boyd}},\
  }\bibfield  {title} {\bibinfo {title} {Supersensitive measurement of angular
  displacements using entangled photons},\ }\href
  {https://doi.org/10.1103/PhysRevA.83.053829} {\bibfield  {journal} {\bibinfo
  {journal} {Phys. Rev. A}\ }\textbf {\bibinfo {volume} {83}},\ \bibinfo
  {pages} {053829} (\bibinfo {year} {2011}{\natexlab{a}})}\BibitemShut
  {NoStop}%
\bibitem [{\citenamefont {Luo}\ \emph {et~al.}(2019)\citenamefont {Luo},
  \citenamefont {Zhong}, \citenamefont {Erhard}, \citenamefont {Wang},
  \citenamefont {Peng}, \citenamefont {Krenn}, \citenamefont {Jiang},
  \citenamefont {Li}, \citenamefont {Liu}, \citenamefont {Lu} \emph
  {et~al.}}]{luo2019prl}%
  \BibitemOpen
  \bibfield  {author} {\bibinfo {author} {\bibfnamefont {Y.-H.}\ \bibnamefont
  {Luo}}, \bibinfo {author} {\bibfnamefont {H.-S.}\ \bibnamefont {Zhong}},
  \bibinfo {author} {\bibfnamefont {M.}~\bibnamefont {Erhard}}, \bibinfo
  {author} {\bibfnamefont {X.-L.}\ \bibnamefont {Wang}}, \bibinfo {author}
  {\bibfnamefont {L.-C.}\ \bibnamefont {Peng}}, \bibinfo {author}
  {\bibfnamefont {M.}~\bibnamefont {Krenn}}, \bibinfo {author} {\bibfnamefont
  {X.}~\bibnamefont {Jiang}}, \bibinfo {author} {\bibfnamefont
  {L.}~\bibnamefont {Li}}, \bibinfo {author} {\bibfnamefont {N.-L.}\
  \bibnamefont {Liu}}, \bibinfo {author} {\bibfnamefont {C.-Y.}\ \bibnamefont
  {Lu}}, \emph {et~al.},\ }\bibfield  {title} {\bibinfo {title} {Quantum
  teleportation in high dimensions},\ }\href
  {https://doi.org/10.1103/PhysRevLett.123.070505} {\bibfield  {journal}
  {\bibinfo  {journal} {Phys. Rev. Lett.}\ }\textbf {\bibinfo {volume} {123}},\
  \bibinfo {pages} {070505} (\bibinfo {year} {2019})}\BibitemShut {NoStop}%
\bibitem [{\citenamefont {Zhang}\ \emph {et~al.}(2017)\citenamefont {Zhang},
  \citenamefont {Zhang}, \citenamefont {Li}, \citenamefont {Zhang},
  \citenamefont {Cheng}, \citenamefont {Li},\ and\ \citenamefont
  {Zhang}}]{zhang2017pra}%
  \BibitemOpen
  \bibfield  {author} {\bibinfo {author} {\bibfnamefont {D.}~\bibnamefont
  {Zhang}}, \bibinfo {author} {\bibfnamefont {Y.}~\bibnamefont {Zhang}},
  \bibinfo {author} {\bibfnamefont {X.}~\bibnamefont {Li}}, \bibinfo {author}
  {\bibfnamefont {D.}~\bibnamefont {Zhang}}, \bibinfo {author} {\bibfnamefont
  {L.}~\bibnamefont {Cheng}}, \bibinfo {author} {\bibfnamefont
  {C.}~\bibnamefont {Li}},\ and\ \bibinfo {author} {\bibfnamefont
  {Y.}~\bibnamefont {Zhang}},\ }\bibfield  {title} {\bibinfo {title}
  {Generation of high-dimensional energy-time-entangled photon pairs},\ }\href
  {https://doi.org/10.1103/PhysRevA.96.053849} {\bibfield  {journal} {\bibinfo
  {journal} {Physical Review A}\ }\textbf {\bibinfo {volume} {96}},\ \bibinfo
  {pages} {053849} (\bibinfo {year} {2017})}\BibitemShut {NoStop}%
\bibitem [{\citenamefont {O’Sullivan-Hale}\ \emph {et~al.}(2005)\citenamefont
  {O’Sullivan-Hale}, \citenamefont {Ali~Khan}, \citenamefont {Boyd},\ and\
  \citenamefont {Howell}}]{osullivan2005prl}%
  \BibitemOpen
  \bibfield  {author} {\bibinfo {author} {\bibfnamefont {M.~N.}\ \bibnamefont
  {O’Sullivan-Hale}}, \bibinfo {author} {\bibfnamefont {I.}~\bibnamefont
  {Ali~Khan}}, \bibinfo {author} {\bibfnamefont {R.~W.}\ \bibnamefont {Boyd}},\
  and\ \bibinfo {author} {\bibfnamefont {J.~C.}\ \bibnamefont {Howell}},\
  }\bibfield  {title} {\bibinfo {title} {Pixel entanglement: experimental
  realization of optically entangled d= 3 and d= 6 qudits},\ }\href
  {https://doi.org/10.1103/PhysRevLett.94.220501} {\bibfield  {journal}
  {\bibinfo  {journal} {Physical review letters}\ }\textbf {\bibinfo {volume}
  {94}},\ \bibinfo {pages} {220501} (\bibinfo {year} {2005})}\BibitemShut
  {NoStop}%
\bibitem [{\citenamefont {Yao}\ and\ \citenamefont
  {Padgett}(2011)}]{yao2011advoptphot}%
  \BibitemOpen
  \bibfield  {author} {\bibinfo {author} {\bibfnamefont {A.~M.}\ \bibnamefont
  {Yao}}\ and\ \bibinfo {author} {\bibfnamefont {M.~J.}\ \bibnamefont
  {Padgett}},\ }\bibfield  {title} {\bibinfo {title} {Orbital angular momentum:
  origins, behavior and applications},\ }\href
  {https://doi.org/10.1364/AOP.3.000161} {\bibfield  {journal} {\bibinfo
  {journal} {Advances in optics and photonics}\ }\textbf {\bibinfo {volume}
  {3}},\ \bibinfo {pages} {161} (\bibinfo {year} {2011})}\BibitemShut {NoStop}%
\bibitem [{\citenamefont {Allen}\ \emph {et~al.}(1992)\citenamefont {Allen},
  \citenamefont {Beijersbergen}, \citenamefont {Spreeuw},\ and\ \citenamefont
  {Woerdman}}]{allen1992pra}%
  \BibitemOpen
  \bibfield  {author} {\bibinfo {author} {\bibfnamefont {L.}~\bibnamefont
  {Allen}}, \bibinfo {author} {\bibfnamefont {M.~W.}\ \bibnamefont
  {Beijersbergen}}, \bibinfo {author} {\bibfnamefont {R.~J.~C.}\ \bibnamefont
  {Spreeuw}},\ and\ \bibinfo {author} {\bibfnamefont {J.~P.}\ \bibnamefont
  {Woerdman}},\ }\bibfield  {title} {\bibinfo {title} {Orbital angular momentum
  of light and the transformation of laguerre-gaussian laser modes},\ }\href
  {https://doi.org/10.1103/PhysRevA.45.8185} {\bibfield  {journal} {\bibinfo
  {journal} {Phys. Rev. A}\ }\textbf {\bibinfo {volume} {45}},\ \bibinfo
  {pages} {8185} (\bibinfo {year} {1992})}\BibitemShut {NoStop}%
\bibitem [{\citenamefont {Olvera-Santamar{\'\i}a}\ and\ \citenamefont
  {Ostrovsky}(2023)}]{olvera2023jopt}%
  \BibitemOpen
  \bibfield  {author} {\bibinfo {author} {\bibfnamefont {M.~{\'A}.}\
  \bibnamefont {Olvera-Santamar{\'\i}a}}\ and\ \bibinfo {author} {\bibfnamefont
  {A.}~\bibnamefont {Ostrovsky}},\ }\bibfield  {title} {\bibinfo {title}
  {Synthesis of partially coherent bessel-mode vortex-beams with radial
  coherence},\ }\href {https://doi.org/10.1088/2040-8986/aceb15} {\bibfield
  {journal} {\bibinfo  {journal} {Journal of Optics}\ }\textbf {\bibinfo
  {volume} {25}},\ \bibinfo {pages} {095601} (\bibinfo {year}
  {2023})}\BibitemShut {NoStop}%
\bibitem [{\citenamefont {Wang}\ \emph {et~al.}(2024)\citenamefont {Wang},
  \citenamefont {Bai}, \citenamefont {Xie}, \citenamefont {Huang},\ and\
  \citenamefont {Guo}}]{wang2024optcomm}%
  \BibitemOpen
  \bibfield  {author} {\bibinfo {author} {\bibfnamefont {Y.}~\bibnamefont
  {Wang}}, \bibinfo {author} {\bibfnamefont {L.}~\bibnamefont {Bai}}, \bibinfo
  {author} {\bibfnamefont {J.}~\bibnamefont {Xie}}, \bibinfo {author}
  {\bibfnamefont {C.}~\bibnamefont {Huang}},\ and\ \bibinfo {author}
  {\bibfnamefont {L.}~\bibnamefont {Guo}},\ }\bibfield  {title} {\bibinfo
  {title} {Radial spectrum spread of laguerre-gaussian beam transmission in
  weak compressible turbulence},\ }\href
  {https://doi.org/10.1016/j.optcom.2023.130111} {\bibfield  {journal}
  {\bibinfo  {journal} {Optics Communications}\ }\textbf {\bibinfo {volume}
  {554}},\ \bibinfo {pages} {130111} (\bibinfo {year} {2024})}\BibitemShut
  {NoStop}%
\bibitem [{\citenamefont {Zhang}\ \emph {et~al.}(2018)\citenamefont {Zhang},
  \citenamefont {Qiu}, \citenamefont {Zhang},\ and\ \citenamefont
  {Chen}}]{zhang2018pra}%
  \BibitemOpen
  \bibfield  {author} {\bibinfo {author} {\bibfnamefont {D.}~\bibnamefont
  {Zhang}}, \bibinfo {author} {\bibfnamefont {X.}~\bibnamefont {Qiu}}, \bibinfo
  {author} {\bibfnamefont {W.}~\bibnamefont {Zhang}},\ and\ \bibinfo {author}
  {\bibfnamefont {L.}~\bibnamefont {Chen}},\ }\bibfield  {title} {\bibinfo
  {title} {Violation of a bell inequality in two-dimensional state spaces for
  radial quantum number},\ }\href {https://doi.org/10.1103/PhysRevA.98.042134}
  {\bibfield  {journal} {\bibinfo  {journal} {Physical Review A}\ }\textbf
  {\bibinfo {volume} {98}},\ \bibinfo {pages} {042134} (\bibinfo {year}
  {2018})}\BibitemShut {NoStop}%
\bibitem [{\citenamefont {Zhang}\ \emph {et~al.}(2022)\citenamefont {Zhang},
  \citenamefont {Zhang}, \citenamefont {Qiu}, \citenamefont {Chen},
  \citenamefont {Franke-Arnold},\ and\ \citenamefont
  {Chen}}]{zhang2022photres}%
  \BibitemOpen
  \bibfield  {author} {\bibinfo {author} {\bibfnamefont {Z.}~\bibnamefont
  {Zhang}}, \bibinfo {author} {\bibfnamefont {D.}~\bibnamefont {Zhang}},
  \bibinfo {author} {\bibfnamefont {X.}~\bibnamefont {Qiu}}, \bibinfo {author}
  {\bibfnamefont {Y.}~\bibnamefont {Chen}}, \bibinfo {author} {\bibfnamefont
  {S.}~\bibnamefont {Franke-Arnold}},\ and\ \bibinfo {author} {\bibfnamefont
  {L.}~\bibnamefont {Chen}},\ }\bibfield  {title} {\bibinfo {title}
  {Experimental investigation of the uncertainty principle for radial degrees
  of freedom},\ }\href {https://doi.org/10.1364/PRJ.443691} {\bibfield
  {journal} {\bibinfo  {journal} {Photonics Research}\ }\textbf {\bibinfo
  {volume} {10}},\ \bibinfo {pages} {2223} (\bibinfo {year}
  {2022})}\BibitemShut {NoStop}%
\bibitem [{\citenamefont {Chen}\ \emph {et~al.}(2019)\citenamefont {Chen},
  \citenamefont {Ma}, \citenamefont {Qiu}, \citenamefont {Zhang}, \citenamefont
  {Zhang},\ and\ \citenamefont {Boyd}}]{chen2019prl}%
  \BibitemOpen
  \bibfield  {author} {\bibinfo {author} {\bibfnamefont {L.}~\bibnamefont
  {Chen}}, \bibinfo {author} {\bibfnamefont {T.}~\bibnamefont {Ma}}, \bibinfo
  {author} {\bibfnamefont {X.}~\bibnamefont {Qiu}}, \bibinfo {author}
  {\bibfnamefont {D.}~\bibnamefont {Zhang}}, \bibinfo {author} {\bibfnamefont
  {W.}~\bibnamefont {Zhang}},\ and\ \bibinfo {author} {\bibfnamefont {R.~W.}\
  \bibnamefont {Boyd}},\ }\bibfield  {title} {\bibinfo {title} {Realization of
  the einstein-podolsky-rosen paradox using radial position and radial momentum
  variables},\ }\href {https://doi.org/10.1103/PhysRevLett.123.060403}
  {\bibfield  {journal} {\bibinfo  {journal} {Phys. Rev. Lett.}\ }\textbf
  {\bibinfo {volume} {123}},\ \bibinfo {pages} {060403} (\bibinfo {year}
  {2019})}\BibitemShut {NoStop}%
\bibitem [{\citenamefont {Ekert}\ and\ \citenamefont
  {Knight}(1995)}]{ekert1995ajp}%
  \BibitemOpen
  \bibfield  {author} {\bibinfo {author} {\bibfnamefont {A.}~\bibnamefont
  {Ekert}}\ and\ \bibinfo {author} {\bibfnamefont {P.~L.}\ \bibnamefont
  {Knight}},\ }\bibfield  {title} {\bibinfo {title} {Entangled quantum systems
  and the schmidt decomposition},\ }\href {https://doi.org/10.1119/1.17904}
  {\bibfield  {journal} {\bibinfo  {journal} {American Journal of Physics}\
  }\textbf {\bibinfo {volume} {63}},\ \bibinfo {pages} {415} (\bibinfo {year}
  {1995})}\BibitemShut {NoStop}%
\bibitem [{\citenamefont {Mirhosseini}\ \emph {et~al.}(2013)\citenamefont
  {Mirhosseini}, \citenamefont {Malik}, \citenamefont {Shi},\ and\
  \citenamefont {Boyd}}]{mirhosseini2013natcomm}%
  \BibitemOpen
  \bibfield  {author} {\bibinfo {author} {\bibfnamefont {M.}~\bibnamefont
  {Mirhosseini}}, \bibinfo {author} {\bibfnamefont {M.}~\bibnamefont {Malik}},
  \bibinfo {author} {\bibfnamefont {Z.}~\bibnamefont {Shi}},\ and\ \bibinfo
  {author} {\bibfnamefont {R.~W.}\ \bibnamefont {Boyd}},\ }\bibfield  {title}
  {\bibinfo {title} {Efficient separation of the orbital angular momentum
  eigenstates of light},\ }\href {https://doi.org/10.1038/ncomms3781}
  {\bibfield  {journal} {\bibinfo  {journal} {Nature Communications}\ }\textbf
  {\bibinfo {volume} {4}},\ \bibinfo {pages} {2781} (\bibinfo {year}
  {2013})}\BibitemShut {NoStop}%
\bibitem [{\citenamefont {Berkhout}\ \emph {et~al.}(2010)\citenamefont
  {Berkhout}, \citenamefont {Lavery}, \citenamefont {Courtial}, \citenamefont
  {Beijersbergen},\ and\ \citenamefont {Padgett}}]{berkhout2010prl}%
  \BibitemOpen
  \bibfield  {author} {\bibinfo {author} {\bibfnamefont {G.~C.~G.}\
  \bibnamefont {Berkhout}}, \bibinfo {author} {\bibfnamefont {M.~P.~J.}\
  \bibnamefont {Lavery}}, \bibinfo {author} {\bibfnamefont {J.}~\bibnamefont
  {Courtial}}, \bibinfo {author} {\bibfnamefont {M.~W.}\ \bibnamefont
  {Beijersbergen}},\ and\ \bibinfo {author} {\bibfnamefont {M.~J.}\
  \bibnamefont {Padgett}},\ }\bibfield  {title} {\bibinfo {title} {Efficient
  {{Sorting}} of {{Orbital Angular Momentum States}} of {{Light}}},\ }\href
  {https://doi.org/10.1103/PhysRevLett.105.153601} {\bibfield  {journal}
  {\bibinfo  {journal} {Physical Review Letters}\ }\textbf {\bibinfo {volume}
  {105}},\ \bibinfo {pages} {153601} (\bibinfo {year} {2010})}\BibitemShut
  {NoStop}%
\bibitem [{\citenamefont {Sahu}\ \emph {et~al.}(2018)\citenamefont {Sahu},
  \citenamefont {Chaudhary}, \citenamefont {Khare}, \citenamefont
  {Bhattacharya}, \citenamefont {Wanare},\ and\ \citenamefont
  {Jha}}]{sahu2018optexp}%
  \BibitemOpen
  \bibfield  {author} {\bibinfo {author} {\bibfnamefont {R.}~\bibnamefont
  {Sahu}}, \bibinfo {author} {\bibfnamefont {S.}~\bibnamefont {Chaudhary}},
  \bibinfo {author} {\bibfnamefont {K.}~\bibnamefont {Khare}}, \bibinfo
  {author} {\bibfnamefont {M.}~\bibnamefont {Bhattacharya}}, \bibinfo {author}
  {\bibfnamefont {H.}~\bibnamefont {Wanare}},\ and\ \bibinfo {author}
  {\bibfnamefont {A.~K.}\ \bibnamefont {Jha}},\ }\bibfield  {title} {\bibinfo
  {title} {Angular lens},\ }\href {https://doi.org/10.1364/oe.26.008709}
  {\bibfield  {journal} {\bibinfo  {journal} {Optics Express}\ }\textbf
  {\bibinfo {volume} {26}},\ \bibinfo {pages} {8709} (\bibinfo {year}
  {2018})}\BibitemShut {NoStop}%
\bibitem [{\citenamefont {Mair}\ \emph {et~al.}(2001)\citenamefont {Mair},
  \citenamefont {Vaziri}, \citenamefont {Weihs},\ and\ \citenamefont
  {Zeilinger}}]{mair2001nature}%
  \BibitemOpen
  \bibfield  {author} {\bibinfo {author} {\bibfnamefont {A.}~\bibnamefont
  {Mair}}, \bibinfo {author} {\bibfnamefont {A.}~\bibnamefont {Vaziri}},
  \bibinfo {author} {\bibfnamefont {G.}~\bibnamefont {Weihs}},\ and\ \bibinfo
  {author} {\bibfnamefont {A.}~\bibnamefont {Zeilinger}},\ }\bibfield  {title}
  {\bibinfo {title} {Entanglement of the orbital angular momentum states of
  photons},\ }\href {https://doi.org/10.1038/35085529} {\bibfield  {journal}
  {\bibinfo  {journal} {Nature}\ }\textbf {\bibinfo {volume} {412}},\ \bibinfo
  {pages} {313} (\bibinfo {year} {2001})}\BibitemShut {NoStop}%
\bibitem [{\citenamefont {Heckenberg}\ \emph {et~al.}(1992)\citenamefont
  {Heckenberg}, \citenamefont {McDuff}, \citenamefont {Smith},\ and\
  \citenamefont {White}}]{heckenberg1992optlett}%
  \BibitemOpen
  \bibfield  {author} {\bibinfo {author} {\bibfnamefont {N.}~\bibnamefont
  {Heckenberg}}, \bibinfo {author} {\bibfnamefont {R.}~\bibnamefont {McDuff}},
  \bibinfo {author} {\bibfnamefont {C.}~\bibnamefont {Smith}},\ and\ \bibinfo
  {author} {\bibfnamefont {A.}~\bibnamefont {White}},\ }\bibfield  {title}
  {\bibinfo {title} {Generation of optical phase singularities by
  computer-generated holograms},\ }\href {https://doi.org/10.1364/OL.17.000221}
  {\bibfield  {journal} {\bibinfo  {journal} {Optics letters}\ }\textbf
  {\bibinfo {volume} {17}},\ \bibinfo {pages} {221} (\bibinfo {year}
  {1992})}\BibitemShut {NoStop}%
\bibitem [{\citenamefont {Jha}\ \emph {et~al.}(2011{\natexlab{b}})\citenamefont
  {Jha}, \citenamefont {Agarwal},\ and\ \citenamefont {Boyd}}]{jha2011pra}%
  \BibitemOpen
  \bibfield  {author} {\bibinfo {author} {\bibfnamefont {A.~K.}\ \bibnamefont
  {Jha}}, \bibinfo {author} {\bibfnamefont {G.~S.}\ \bibnamefont {Agarwal}},\
  and\ \bibinfo {author} {\bibfnamefont {R.~W.}\ \bibnamefont {Boyd}},\
  }\bibfield  {title} {\bibinfo {title} {Partial angular coherence and the
  angular schmidt spectrum of entangled two-photon fields},\ }\href
  {https://doi.org/10.1103/PhysRevA.84.063847} {\bibfield  {journal} {\bibinfo
  {journal} {Physical Review A}\ }\textbf {\bibinfo {volume} {84}},\ \bibinfo
  {pages} {063847} (\bibinfo {year} {2011}{\natexlab{b}})}\BibitemShut
  {NoStop}%
\bibitem [{\citenamefont {Di~Lorenzo~Pires}\ \emph {et~al.}(2010)\citenamefont
  {Di~Lorenzo~Pires}, \citenamefont {Florijn},\ and\ \citenamefont {van
  Exter}}]{pires2010prl}%
  \BibitemOpen
  \bibfield  {author} {\bibinfo {author} {\bibfnamefont {H.}~\bibnamefont
  {Di~Lorenzo~Pires}}, \bibinfo {author} {\bibfnamefont {H.~C.~B.}\
  \bibnamefont {Florijn}},\ and\ \bibinfo {author} {\bibfnamefont {M.~P.}\
  \bibnamefont {van Exter}},\ }\bibfield  {title} {\bibinfo {title}
  {Measurement of the spiral spectrum of entangled two-photon states},\ }\href
  {https://doi.org/10.1103/PhysRevLett.104.020505} {\bibfield  {journal}
  {\bibinfo  {journal} {Phys. Rev. Lett.}\ }\textbf {\bibinfo {volume} {104}},\
  \bibinfo {pages} {020505} (\bibinfo {year} {2010})}\BibitemShut {NoStop}%
\bibitem [{\citenamefont {Kulkarni}\ \emph {et~al.}(2017)\citenamefont
  {Kulkarni}, \citenamefont {Sahu}, \citenamefont {Maga{\~n}a-Loaiza},
  \citenamefont {Boyd},\ and\ \citenamefont {Jha}}]{kulkarni2017natcomm}%
  \BibitemOpen
  \bibfield  {author} {\bibinfo {author} {\bibfnamefont {G.}~\bibnamefont
  {Kulkarni}}, \bibinfo {author} {\bibfnamefont {R.}~\bibnamefont {Sahu}},
  \bibinfo {author} {\bibfnamefont {O.~S.}\ \bibnamefont {Maga{\~n}a-Loaiza}},
  \bibinfo {author} {\bibfnamefont {R.~W.}\ \bibnamefont {Boyd}},\ and\
  \bibinfo {author} {\bibfnamefont {A.~K.}\ \bibnamefont {Jha}},\ }\bibfield
  {title} {\bibinfo {title} {Single-shot measurement of the
  orbital-angular-momentum spectrum of light},\ }\href
  {https://doi.org/10.1038/s41467-017-01215-x} {\bibfield  {journal} {\bibinfo
  {journal} {Nat. Commun.}\ }\textbf {\bibinfo {volume} {8}},\ \bibinfo {pages}
  {1054} (\bibinfo {year} {2017})}\BibitemShut {NoStop}%
\bibitem [{\citenamefont {Kulkarni}\ \emph {et~al.}(2018)\citenamefont
  {Kulkarni}, \citenamefont {Taneja}, \citenamefont {Aarav},\ and\
  \citenamefont {Jha}}]{kulkarni2018pra}%
  \BibitemOpen
  \bibfield  {author} {\bibinfo {author} {\bibfnamefont {G.}~\bibnamefont
  {Kulkarni}}, \bibinfo {author} {\bibfnamefont {L.}~\bibnamefont {Taneja}},
  \bibinfo {author} {\bibfnamefont {S.}~\bibnamefont {Aarav}},\ and\ \bibinfo
  {author} {\bibfnamefont {A.~K.}\ \bibnamefont {Jha}},\ }\bibfield  {title}
  {\bibinfo {title} {Angular schmidt spectrum of entangled photons: Derivation
  of an exact formula and experimental characterization for noncollinear phase
  matching},\ }\href {https://doi.org/10.1103/PhysRevA.97.063846} {\bibfield
  {journal} {\bibinfo  {journal} {Physical Review A}\ }\textbf {\bibinfo
  {volume} {97}},\ \bibinfo {pages} {063846} (\bibinfo {year}
  {2018})}\BibitemShut {NoStop}%
\bibitem [{\citenamefont {Karan}\ \emph {et~al.}(2025)\citenamefont {Karan},
  \citenamefont {Van~Exter},\ and\ \citenamefont {Jha}}]{karan2025sciadv}%
  \BibitemOpen
  \bibfield  {author} {\bibinfo {author} {\bibfnamefont {S.}~\bibnamefont
  {Karan}}, \bibinfo {author} {\bibfnamefont {M.~P.}\ \bibnamefont
  {Van~Exter}},\ and\ \bibinfo {author} {\bibfnamefont {A.~K.}\ \bibnamefont
  {Jha}},\ }\bibfield  {title} {\bibinfo {title} {Broadband uniform-efficiency
  oam-mode detector},\ }\href {https://doi.org/10.1126/sciadv.adq7201}
  {\bibfield  {journal} {\bibinfo  {journal} {Science Advances}\ }\textbf
  {\bibinfo {volume} {11}},\ \bibinfo {pages} {eadq7201} (\bibinfo {year}
  {2025})}\BibitemShut {NoStop}%
\bibitem [{\citenamefont {Zhou}\ \emph {et~al.}(2017)\citenamefont {Zhou},
  \citenamefont {Mirhosseini}, \citenamefont {Fu}, \citenamefont {Zhao},
  \citenamefont {Hashemi~Rafsanjani}, \citenamefont {Willner},\ and\
  \citenamefont {Boyd}}]{zhou2017prl}%
  \BibitemOpen
  \bibfield  {author} {\bibinfo {author} {\bibfnamefont {Y.}~\bibnamefont
  {Zhou}}, \bibinfo {author} {\bibfnamefont {M.}~\bibnamefont {Mirhosseini}},
  \bibinfo {author} {\bibfnamefont {D.}~\bibnamefont {Fu}}, \bibinfo {author}
  {\bibfnamefont {J.}~\bibnamefont {Zhao}}, \bibinfo {author} {\bibfnamefont
  {S.~M.}\ \bibnamefont {Hashemi~Rafsanjani}}, \bibinfo {author} {\bibfnamefont
  {A.~E.}\ \bibnamefont {Willner}},\ and\ \bibinfo {author} {\bibfnamefont
  {R.~W.}\ \bibnamefont {Boyd}},\ }\bibfield  {title} {\bibinfo {title}
  {Sorting photons by radial quantum number},\ }\href
  {https://doi.org/10.1103/PhysRevLett.119.263602} {\bibfield  {journal}
  {\bibinfo  {journal} {Phys. Rev. Lett.}\ }\textbf {\bibinfo {volume} {119}},\
  \bibinfo {pages} {263602} (\bibinfo {year} {2017})}\BibitemShut {NoStop}%
\bibitem [{\citenamefont {Fu}\ \emph {et~al.}(2018)\citenamefont {Fu},
  \citenamefont {Zhou}, \citenamefont {Qi}, \citenamefont {Oliver},
  \citenamefont {Wang}, \citenamefont {Rafsanjani}, \citenamefont {Zhao},
  \citenamefont {Mirhosseini}, \citenamefont {Shi}, \citenamefont {Zhang},\
  and\ \citenamefont {Boyd}}]{fu2018optexp}%
  \BibitemOpen
  \bibfield  {author} {\bibinfo {author} {\bibfnamefont {D.}~\bibnamefont
  {Fu}}, \bibinfo {author} {\bibfnamefont {Y.}~\bibnamefont {Zhou}}, \bibinfo
  {author} {\bibfnamefont {R.}~\bibnamefont {Qi}}, \bibinfo {author}
  {\bibfnamefont {S.}~\bibnamefont {Oliver}}, \bibinfo {author} {\bibfnamefont
  {Y.}~\bibnamefont {Wang}}, \bibinfo {author} {\bibfnamefont {S.~M.~H.}\
  \bibnamefont {Rafsanjani}}, \bibinfo {author} {\bibfnamefont
  {J.}~\bibnamefont {Zhao}}, \bibinfo {author} {\bibfnamefont {M.}~\bibnamefont
  {Mirhosseini}}, \bibinfo {author} {\bibfnamefont {Z.}~\bibnamefont {Shi}},
  \bibinfo {author} {\bibfnamefont {P.}~\bibnamefont {Zhang}},\ and\ \bibinfo
  {author} {\bibfnamefont {R.~W.}\ \bibnamefont {Boyd}},\ }\bibfield  {title}
  {\bibinfo {title} {Realization of a scalable laguerre\&\#x02013;gaussian mode
  sorter based on a robust radial mode sorter},\ }\href
  {https://doi.org/10.1364/OE.26.033057} {\bibfield  {journal} {\bibinfo
  {journal} {Opt. Express}\ }\textbf {\bibinfo {volume} {26}},\ \bibinfo
  {pages} {33057} (\bibinfo {year} {2018})}\BibitemShut {NoStop}%
\bibitem [{\citenamefont {Choudhary}\ \emph {et~al.}(2018)\citenamefont
  {Choudhary}, \citenamefont {Sampson}, \citenamefont {Miyamoto}, \citenamefont
  {Maga{\~n}a-Loaiza}, \citenamefont {Rafsanjani}, \citenamefont
  {Mirhosseini},\ and\ \citenamefont {Boyd}}]{choudhary2018optlett}%
  \BibitemOpen
  \bibfield  {author} {\bibinfo {author} {\bibfnamefont {S.}~\bibnamefont
  {Choudhary}}, \bibinfo {author} {\bibfnamefont {R.}~\bibnamefont {Sampson}},
  \bibinfo {author} {\bibfnamefont {Y.}~\bibnamefont {Miyamoto}}, \bibinfo
  {author} {\bibfnamefont {O.~S.}\ \bibnamefont {Maga{\~n}a-Loaiza}}, \bibinfo
  {author} {\bibfnamefont {S.~M.~H.}\ \bibnamefont {Rafsanjani}}, \bibinfo
  {author} {\bibfnamefont {M.}~\bibnamefont {Mirhosseini}},\ and\ \bibinfo
  {author} {\bibfnamefont {R.~W.}\ \bibnamefont {Boyd}},\ }\bibfield  {title}
  {\bibinfo {title} {Measurement of the radial mode spectrum of photons through
  a phase-retrieval method},\ }\href {https://doi.org/10.1364/OL.43.006101}
  {\bibfield  {journal} {\bibinfo  {journal} {Opt. Lett.}\ }\textbf {\bibinfo
  {volume} {43}},\ \bibinfo {pages} {6101} (\bibinfo {year}
  {2018})}\BibitemShut {NoStop}%
\bibitem [{\citenamefont {Bouchard}\ \emph {et~al.}(2018)\citenamefont
  {Bouchard}, \citenamefont {Valencia}, \citenamefont {Brandt}, \citenamefont
  {Fickler}, \citenamefont {Huber},\ and\ \citenamefont
  {Malik}}]{bouchard2018optexp}%
  \BibitemOpen
  \bibfield  {author} {\bibinfo {author} {\bibfnamefont {F.}~\bibnamefont
  {Bouchard}}, \bibinfo {author} {\bibfnamefont {N.~H.}\ \bibnamefont
  {Valencia}}, \bibinfo {author} {\bibfnamefont {F.}~\bibnamefont {Brandt}},
  \bibinfo {author} {\bibfnamefont {R.}~\bibnamefont {Fickler}}, \bibinfo
  {author} {\bibfnamefont {M.}~\bibnamefont {Huber}},\ and\ \bibinfo {author}
  {\bibfnamefont {M.}~\bibnamefont {Malik}},\ }\bibfield  {title} {\bibinfo
  {title} {Measuring azimuthal and radial modes of photons},\ }\href
  {https://doi.org/10.1364/OE.26.031925} {\bibfield  {journal} {\bibinfo
  {journal} {Optics Express}\ }\textbf {\bibinfo {volume} {26}},\ \bibinfo
  {pages} {31925} (\bibinfo {year} {2018})}\BibitemShut {NoStop}%
\bibitem [{\citenamefont {Offer}\ \emph {et~al.}(2018)\citenamefont {Offer},
  \citenamefont {Stulga}, \citenamefont {Riis}, \citenamefont {Franke-Arnold},\
  and\ \citenamefont {Arnold}}]{offer2018commphy}%
  \BibitemOpen
  \bibfield  {author} {\bibinfo {author} {\bibfnamefont {R.~F.}\ \bibnamefont
  {Offer}}, \bibinfo {author} {\bibfnamefont {D.}~\bibnamefont {Stulga}},
  \bibinfo {author} {\bibfnamefont {E.}~\bibnamefont {Riis}}, \bibinfo {author}
  {\bibfnamefont {S.}~\bibnamefont {Franke-Arnold}},\ and\ \bibinfo {author}
  {\bibfnamefont {A.~S.}\ \bibnamefont {Arnold}},\ }\bibfield  {title}
  {\bibinfo {title} {Spiral bandwidth of four-wave mixing in rb vapour},\
  }\href {https://doi.org/10.1038/s42005-018-0077-5} {\bibfield  {journal}
  {\bibinfo  {journal} {Communications Physics}\ }\textbf {\bibinfo {volume}
  {1}},\ \bibinfo {pages} {84} (\bibinfo {year} {2018})}\BibitemShut {NoStop}%
\bibitem [{\citenamefont {Schneeloch}\ and\ \citenamefont
  {Howell}(2016)}]{schneeloch2016jopt}%
  \BibitemOpen
  \bibfield  {author} {\bibinfo {author} {\bibfnamefont {J.}~\bibnamefont
  {Schneeloch}}\ and\ \bibinfo {author} {\bibfnamefont {J.~C.}\ \bibnamefont
  {Howell}},\ }\bibfield  {title} {\bibinfo {title} {Introduction to the
  transverse spatial correlations in spontaneous parametric down-conversion
  through the biphoton birth zone},\ }\href
  {https://doi.org/10.1088/2040-8978/18/5/053501} {\bibfield  {journal}
  {\bibinfo  {journal} {Journal of Optics}\ }\textbf {\bibinfo {volume} {18}},\
  \bibinfo {pages} {053501} (\bibinfo {year} {2016})}\BibitemShut {NoStop}%
\bibitem [{\citenamefont {Karan}\ \emph {et~al.}(2020)\citenamefont {Karan},
  \citenamefont {Aarav}, \citenamefont {Bharadhwaj}, \citenamefont {Taneja},
  \citenamefont {De}, \citenamefont {Kulkarni}, \citenamefont {Meher},\ and\
  \citenamefont {Jha}}]{karan2020jopt}%
  \BibitemOpen
  \bibfield  {author} {\bibinfo {author} {\bibfnamefont {S.}~\bibnamefont
  {Karan}}, \bibinfo {author} {\bibfnamefont {S.}~\bibnamefont {Aarav}},
  \bibinfo {author} {\bibfnamefont {H.}~\bibnamefont {Bharadhwaj}}, \bibinfo
  {author} {\bibfnamefont {L.}~\bibnamefont {Taneja}}, \bibinfo {author}
  {\bibfnamefont {A.}~\bibnamefont {De}}, \bibinfo {author} {\bibfnamefont
  {G.}~\bibnamefont {Kulkarni}}, \bibinfo {author} {\bibfnamefont
  {N.}~\bibnamefont {Meher}},\ and\ \bibinfo {author} {\bibfnamefont {A.~K.}\
  \bibnamefont {Jha}},\ }\bibfield  {title} {\bibinfo {title} {Phase matching
  in $\beta$-barium borate crystals for spontaneous parametric
  down-conversion},\ }\href {https://doi.org/10.1088/2040-8986/ab89e4}
  {\bibfield  {journal} {\bibinfo  {journal} {Journal of Optics}\ }\textbf
  {\bibinfo {volume} {22}},\ \bibinfo {pages} {083501} (\bibinfo {year}
  {2020})}\BibitemShut {NoStop}%
\bibitem [{\citenamefont {Walborn}\ \emph {et~al.}(2010)\citenamefont
  {Walborn}, \citenamefont {Monken}, \citenamefont {P{\'a}dua},\ and\
  \citenamefont {Ribeiro}}]{walborn2010phyrep}%
  \BibitemOpen
  \bibfield  {author} {\bibinfo {author} {\bibfnamefont {S.~P.}\ \bibnamefont
  {Walborn}}, \bibinfo {author} {\bibfnamefont {C.}~\bibnamefont {Monken}},
  \bibinfo {author} {\bibfnamefont {S.}~\bibnamefont {P{\'a}dua}},\ and\
  \bibinfo {author} {\bibfnamefont {P.~S.}\ \bibnamefont {Ribeiro}},\
  }\bibfield  {title} {\bibinfo {title} {Spatial correlations in parametric
  down-conversion},\ }\href {https://doi.org/10.1016/j.physrep.2010.06.003}
  {\bibfield  {journal} {\bibinfo  {journal} {Physics Reports}\ }\textbf
  {\bibinfo {volume} {495}},\ \bibinfo {pages} {87} (\bibinfo {year}
  {2010})}\BibitemShut {NoStop}%
\bibitem [{\citenamefont {Karan}\ \emph {et~al.}(2023)\citenamefont {Karan},
  \citenamefont {Prasad},\ and\ \citenamefont {Jha}}]{karan2023prapp}%
  \BibitemOpen
  \bibfield  {author} {\bibinfo {author} {\bibfnamefont {S.}~\bibnamefont
  {Karan}}, \bibinfo {author} {\bibfnamefont {R.}~\bibnamefont {Prasad}},\ and\
  \bibinfo {author} {\bibfnamefont {A.~K.}\ \bibnamefont {Jha}},\ }\bibfield
  {title} {\bibinfo {title} {Postselection-free controlled generation of a
  high-dimensional orbital-angular-momentum entangled state},\ }\href
  {https://doi.org/10.1103/PhysRevApplied.20.054027} {\bibfield  {journal}
  {\bibinfo  {journal} {Physical Review Applied}\ }\textbf {\bibinfo {volume}
  {20}},\ \bibinfo {pages} {054027} (\bibinfo {year} {2023})}\BibitemShut
  {NoStop}%
\bibitem [{\citenamefont {Patoary}\ \emph {et~al.}(2019)\citenamefont
  {Patoary}, \citenamefont {Kulkarni},\ and\ \citenamefont
  {Jha}}]{patoary2019josab}%
  \BibitemOpen
  \bibfield  {author} {\bibinfo {author} {\bibfnamefont {A.~S.~M.}\
  \bibnamefont {Patoary}}, \bibinfo {author} {\bibfnamefont {G.}~\bibnamefont
  {Kulkarni}},\ and\ \bibinfo {author} {\bibfnamefont {A.~K.}\ \bibnamefont
  {Jha}},\ }\bibfield  {title} {\bibinfo {title} {Intrinsic degree of coherence
  of classical and quantum states},\ }\href
  {https://doi.org/10.1364/JOSAB.36.002765} {\bibfield  {journal} {\bibinfo
  {journal} {J. Opt. Soc. Am. B}\ }\textbf {\bibinfo {volume} {36}},\ \bibinfo
  {pages} {2765} (\bibinfo {year} {2019})}\BibitemShut {NoStop}%
\bibitem [{\citenamefont {Born}\ and\ \citenamefont
  {Wolf}(1999)}]{born&wolf1999}%
  \BibitemOpen
  \bibfield  {author} {\bibinfo {author} {\bibfnamefont {M.}~\bibnamefont
  {Born}}\ and\ \bibinfo {author} {\bibfnamefont {E.}~\bibnamefont {Wolf}},\
  }\href@noop {} {\emph {\bibinfo {title} {Principles of Optics}}},\ \bibinfo
  {edition} {7th}\ ed.\ (\bibinfo  {publisher} {Cambridge University Press},\
  \bibinfo {address} {Cambridge},\ \bibinfo {year} {1999})\BibitemShut
  {NoStop}%
\bibitem [{\citenamefont {Zia}\ \emph {et~al.}(2023)\citenamefont {Zia},
  \citenamefont {Dehghan}, \citenamefont {D’Errico}, \citenamefont
  {Sciarrino},\ and\ \citenamefont {Karimi}}]{zia2023natphot}%
  \BibitemOpen
  \bibfield  {author} {\bibinfo {author} {\bibfnamefont {D.}~\bibnamefont
  {Zia}}, \bibinfo {author} {\bibfnamefont {N.}~\bibnamefont {Dehghan}},
  \bibinfo {author} {\bibfnamefont {A.}~\bibnamefont {D’Errico}}, \bibinfo
  {author} {\bibfnamefont {F.}~\bibnamefont {Sciarrino}},\ and\ \bibinfo
  {author} {\bibfnamefont {E.}~\bibnamefont {Karimi}},\ }\bibfield  {title}
  {\bibinfo {title} {Interferometric imaging of amplitude and phase of spatial
  biphoton states},\ }\href {https://doi.org/10.1038/s41566-023-01272-3}
  {\bibfield  {journal} {\bibinfo  {journal} {Nature Photonics}\ }\textbf
  {\bibinfo {volume} {17}},\ \bibinfo {pages} {1009} (\bibinfo {year}
  {2023})}\BibitemShut {NoStop}%
\bibitem [{\citenamefont {Jha}\ and\ \citenamefont {Boyd}(2010)}]{jha2010pra}%
  \BibitemOpen
  \bibfield  {author} {\bibinfo {author} {\bibfnamefont {A.~K.}\ \bibnamefont
  {Jha}}\ and\ \bibinfo {author} {\bibfnamefont {R.~W.}\ \bibnamefont {Boyd}},\
  }\bibfield  {title} {\bibinfo {title} {Spatial two-photon coherence of the
  entangled field produced by down-conversion using a partially spatially
  coherent pump beam},\ }\href {https://doi.org/10.1103/PhysRevA.81.013828}
  {\bibfield  {journal} {\bibinfo  {journal} {Phys. Rev. A}\ }\textbf {\bibinfo
  {volume} {81}},\ \bibinfo {pages} {013828} (\bibinfo {year}
  {2010})}\BibitemShut {NoStop}%
\bibitem [{\citenamefont {Meher}\ \emph {et~al.}(2020)\citenamefont {Meher},
  \citenamefont {Patoary}, \citenamefont {Kulkarni},\ and\ \citenamefont
  {Jha}}]{meher2020josab}%
  \BibitemOpen
  \bibfield  {author} {\bibinfo {author} {\bibfnamefont {N.}~\bibnamefont
  {Meher}}, \bibinfo {author} {\bibfnamefont {A.~S.~M.}\ \bibnamefont
  {Patoary}}, \bibinfo {author} {\bibfnamefont {G.}~\bibnamefont {Kulkarni}},\
  and\ \bibinfo {author} {\bibfnamefont {A.~K.}\ \bibnamefont {Jha}},\
  }\bibfield  {title} {\bibinfo {title} {Intrinsic degree of coherence of
  two-qubit states and measures of two-particle quantum correlations},\ }\href
  {https://doi.org/10.1364/JOSAB.384408} {\bibfield  {journal} {\bibinfo
  {journal} {J. Opt. Soc. Am. B}\ }\textbf {\bibinfo {volume} {37}},\ \bibinfo
  {pages} {1224} (\bibinfo {year} {2020})}\BibitemShut {NoStop}%
\bibitem [{\citenamefont {Bhattacharjee}\ \emph {et~al.}(2023)\citenamefont
  {Bhattacharjee}, \citenamefont {Biswas}, \citenamefont {Alonso},\ and\
  \citenamefont {Jha}}]{bhattacharjee2023josaa}%
  \BibitemOpen
  \bibfield  {author} {\bibinfo {author} {\bibfnamefont {A.}~\bibnamefont
  {Bhattacharjee}}, \bibinfo {author} {\bibfnamefont {S.}~\bibnamefont
  {Biswas}}, \bibinfo {author} {\bibfnamefont {M.~A.}\ \bibnamefont {Alonso}},\
  and\ \bibinfo {author} {\bibfnamefont {A.~K.}\ \bibnamefont {Jha}},\
  }\bibfield  {title} {\bibinfo {title} {Coherence in the radial degree of
  freedom},\ }\href {https://doi.org/10.1364/JOSAA.474724} {\bibfield
  {journal} {\bibinfo  {journal} {JOSA A}\ }\textbf {\bibinfo {volume} {40}},\
  \bibinfo {pages} {411} (\bibinfo {year} {2023})}\BibitemShut {NoStop}%
\bibitem [{\citenamefont {Prasad}\ \emph {et~al.}(2025)\citenamefont {Prasad},
  \citenamefont {Senapati}, \citenamefont {Karan}, \citenamefont
  {Bhattacharjee}, \citenamefont {Piccirillo}, \citenamefont {Alonso},\ and\
  \citenamefont {Jha}}]{prasad2025optexp}%
  \BibitemOpen
  \bibfield  {author} {\bibinfo {author} {\bibfnamefont {R.}~\bibnamefont
  {Prasad}}, \bibinfo {author} {\bibfnamefont {N.}~\bibnamefont {Senapati}},
  \bibinfo {author} {\bibfnamefont {S.}~\bibnamefont {Karan}}, \bibinfo
  {author} {\bibfnamefont {A.}~\bibnamefont {Bhattacharjee}}, \bibinfo {author}
  {\bibfnamefont {B.}~\bibnamefont {Piccirillo}}, \bibinfo {author}
  {\bibfnamefont {M.~A.}\ \bibnamefont {Alonso}},\ and\ \bibinfo {author}
  {\bibfnamefont {A.~K.}\ \bibnamefont {Jha}},\ }\bibfield  {title} {\bibinfo
  {title} {Experimental technique for measuring radial coherence},\ }\href
  {https://doi.org/10.1364/OE.555049} {\bibfield  {journal} {\bibinfo
  {journal} {Optics Express}\ }\textbf {\bibinfo {volume} {33}},\ \bibinfo
  {pages} {11693} (\bibinfo {year} {2025})}\BibitemShut {NoStop}%
\bibitem [{\citenamefont {Schneeloch}\ \emph {et~al.}(2019)\citenamefont
  {Schneeloch}, \citenamefont {Tison}, \citenamefont {Fanto}, \citenamefont
  {Alsing},\ and\ \citenamefont {Howland}}]{schneeloch2019natcomm}%
  \BibitemOpen
  \bibfield  {author} {\bibinfo {author} {\bibfnamefont {J.}~\bibnamefont
  {Schneeloch}}, \bibinfo {author} {\bibfnamefont {C.~C.}\ \bibnamefont
  {Tison}}, \bibinfo {author} {\bibfnamefont {M.~L.}\ \bibnamefont {Fanto}},
  \bibinfo {author} {\bibfnamefont {P.~M.}\ \bibnamefont {Alsing}},\ and\
  \bibinfo {author} {\bibfnamefont {G.~A.}\ \bibnamefont {Howland}},\
  }\bibfield  {title} {\bibinfo {title} {Quantifying entanglement in a
  68-billion-dimensional quantum state space},\ }\href
  {https://doi.org/10.1038/s41467-019-10810-z} {\bibfield  {journal} {\bibinfo
  {journal} {Nature Communications}\ }\textbf {\bibinfo {volume} {10}},\
  \bibinfo {pages} {2785} (\bibinfo {year} {2019})}\BibitemShut {NoStop}%
\bibitem [{\citenamefont {Prasad}\ \emph {et~al.}(2024)\citenamefont {Prasad},
  \citenamefont {Wanare}, \citenamefont {Karan}, \citenamefont {Joshi},
  \citenamefont {Bhattacharjee},\ and\ \citenamefont {Jha}}]{prasad2024prapp}%
  \BibitemOpen
  \bibfield  {author} {\bibinfo {author} {\bibfnamefont {R.}~\bibnamefont
  {Prasad}}, \bibinfo {author} {\bibfnamefont {S.}~\bibnamefont {Wanare}},
  \bibinfo {author} {\bibfnamefont {S.}~\bibnamefont {Karan}}, \bibinfo
  {author} {\bibfnamefont {M.~K.}\ \bibnamefont {Joshi}}, \bibinfo {author}
  {\bibfnamefont {A.}~\bibnamefont {Bhattacharjee}},\ and\ \bibinfo {author}
  {\bibfnamefont {A.~K.}\ \bibnamefont {Jha}},\ }\bibfield  {title} {\bibinfo
  {title} {Structured position-momentum-entangled two-photon fields},\ }\href
  {https://doi.org/10.1103/PhysRevApplied.22.064034} {\bibfield  {journal}
  {\bibinfo  {journal} {Physical Review Applied}\ }\textbf {\bibinfo {volume}
  {22}},\ \bibinfo {pages} {064034} (\bibinfo {year} {2024})}\BibitemShut
  {NoStop}%
\bibitem [{\citenamefont {Prevedel}\ \emph {et~al.}(2011)\citenamefont
  {Prevedel}, \citenamefont {Hamel}, \citenamefont {Colbeck}, \citenamefont
  {Fisher},\ and\ \citenamefont {Resch}}]{prevedel2011natphy}%
  \BibitemOpen
  \bibfield  {author} {\bibinfo {author} {\bibfnamefont {R.}~\bibnamefont
  {Prevedel}}, \bibinfo {author} {\bibfnamefont {D.~R.}\ \bibnamefont {Hamel}},
  \bibinfo {author} {\bibfnamefont {R.}~\bibnamefont {Colbeck}}, \bibinfo
  {author} {\bibfnamefont {K.}~\bibnamefont {Fisher}},\ and\ \bibinfo {author}
  {\bibfnamefont {K.~J.}\ \bibnamefont {Resch}},\ }\bibfield  {title} {\bibinfo
  {title} {Experimental investigation of the uncertainty principle in the
  presence of quantum memory and its application to witnessing entanglement},\
  }\href {https://doi.org/10.1038/nphys2048} {\bibfield  {journal} {\bibinfo
  {journal} {Nature Physics}\ }\textbf {\bibinfo {volume} {7}},\ \bibinfo
  {pages} {757} (\bibinfo {year} {2011})}\BibitemShut {NoStop}%
\bibitem [{\citenamefont {Bennett}\ \emph {et~al.}(1996)\citenamefont
  {Bennett}, \citenamefont {Bernstein}, \citenamefont {Popescu},\ and\
  \citenamefont {Schumacher}}]{bennett1996pra}%
  \BibitemOpen
  \bibfield  {author} {\bibinfo {author} {\bibfnamefont {C.~H.}\ \bibnamefont
  {Bennett}}, \bibinfo {author} {\bibfnamefont {H.~J.}\ \bibnamefont
  {Bernstein}}, \bibinfo {author} {\bibfnamefont {S.}~\bibnamefont {Popescu}},\
  and\ \bibinfo {author} {\bibfnamefont {B.}~\bibnamefont {Schumacher}},\
  }\bibfield  {title} {\bibinfo {title} {Concentrating partial entanglement by
  local operations},\ }\href {https://doi.org/10.1103/PhysRevA.53.2046}
  {\bibfield  {journal} {\bibinfo  {journal} {Physical Review A}\ }\textbf
  {\bibinfo {volume} {53}},\ \bibinfo {pages} {2046} (\bibinfo {year}
  {1996})}\BibitemShut {NoStop}%
\end{thebibliography}%

\end{document}